\documentclass[fleqn,usenatbib]{mnras}

\usepackage{newtxtext,newtxmath}

\usepackage[T1]{fontenc}
\usepackage{ae,aecompl}

\usepackage{graphicx}	
\usepackage{amsmath}	
\usepackage{amssymb}	
\usepackage[table,x11names]{xcolor}
\hypersetup{draft}

\newcommand\FeH{\mathrm{[Fe/H]}}
\newcommand\aFe{[\alpha/\mathrm{Fe}]}
\newcommand\MH{\mathrm{[M/H]}}
\newcommand\aM{[\alpha/\mathrm{M}]}
\newcommand\CFe{\mathrm{[C/Fe]}}
\newcommand\NFe{[\mathrm{N/Fe}]}
\newcommand\MgFe{\mathrm{[Mg/Fe]}}
\newcommand\MgMn{\mathrm{[Mg/Mn]}}
\newcommand\AlFe{\mathrm{[Al/Fe]}}
\newcommand\CN{\mathrm{[C/N]}}
\newcommand\CplusNFe{\mathrm{[C+N/Fe]}}
\newcommand\teff{T_{\mathrm{eff}}}

\title[Ages and kinematics of the accreted halo stars]{Ages and kinematics of chemically selected, accreted Milky Way halo stars}

\author[P. Das et al.]{
Payel Das,$^{1}$\thanks{E-mail: payel.das@physics.ox.ac.uk}
Keith Hawkins,$^{2}$
and Paula Jofr\'{e}$^{3}$
\\
$^{1}$Rudolf Peierls Centre for Theoretical Physics, University of Oxford, Parks Road, Oxford, OX1 3PU, UK\\
$^{2}$Department of Astronomy, The University of Texas at Austin, 2515 Speedway, Stop C1400, Austin, Texas, 78712-1205, USA\\
$^{3}$N\'{u}cleo de Astronom\'{i}a, Facultad de Ingenier\'{i}a y Ciencias, Universidad Diego Portales, Ej\'ercito 441, Santiago de Chile}

\date{Accepted XXX. Received YYY; in original form ZZZ}

\pubyear{2019}

\begin{document}
\label{firstpage}
\pagerange{\pageref{firstpage}--\pageref{lastpage}}
\maketitle

\begin{abstract}
We exploit the [Mg/Mn]-[Al/Fe] chemical abundance plane to help identify nearby halo stars in the 14th data release from the APOGEE survey that have been accreted on to the Milky Way. Applying a Gaussian Mixture Model, we find a `blob' of 856 likely accreted stars, with a low disc contamination rate of $\sim$7\%. Cross-matching the sample with the second data release from Gaia gives us access to parallaxes and apparent magnitudes, which place constraints on distances and intrinsic luminosities. Using a Bayesian isochrone pipeline, this enables us to estimate new ages for the accreted stars, with typical uncertainties of $\sim$20\%. Our new catalogue is further supplemented with estimates of orbital parameters.

The blob stars span a metallicities between -0.5 to -2.5, and [Mg/Fe] between -0.1 to 0.5. They constitute $\sim$30\% of the metal-poor ([Fe/H] < -0.8) halo at metallicities of $\sim$-1.4. Our new ages are mainly range between 8 to 13 Gyr, with the oldest stars the metal-poorest, and with the highest [Mg/Fe] abundance. If the blob stars are assumed to belong to a single progenitor, the ages imply that the system merged with our Milky Way around 8 Gyr ago and that star formation proceeded for ~5 Gyr. Dynamical arguments suggest that such a single progenitor would have a total mass of $\sim 10^{11}M_{\odot}$, similar to that found by other authors using chemical evolution models and simulations. Comparing the scatter in the [Mg/Fe]-[Fe/H] plane of the blob stars to that measured for stars belonging to the Large Magellanic Cloud suggests that the blob does indeed contain stars from only one progenitor.

\end{abstract}




\section{Introduction}

The stellar halo contributes about 1\% of the stellar mass budget of the Milky Way, believed to be primarily in the form of old and metal-poor stars. Some of these stars may have formed {\it in situ}, but others arrived from disrupted satellite galaxies and globular clusters \citep{ibata+95,majewski+03,abadi+06,belokurov+07,bell+08,cooper+11}. The engulfed systems disperse in positions and velocities but their ages and metallicities remain bunched \citep{bell+10}. An essential step towards distinguishing between potential origins of stars in the stellar halo is to analyse its chemodynamical and age structure.

The density profile of the stars in the halo, for example, reflects the accumulation of stellar mass there. Star counts of main-sequence turn-off stars \citep{bell+08,sesar+11}, RR Lyrae stars \citep{watkins+09,sesar+13}, BHB stars \citep{deason+11a}, and K giants \citep{kafle+13} suggest a break in the number density around  $R = 15 - 25$ kpc with a power-law index $\alpha\sim2 - 3$ in the inner halo and $\alpha\sim 3.8 - 5$ further out. An examination of BHBs and blue stragglers from SDSS by \cite{deason+14} found evidence that $\alpha\sim6$ beyond $50$ kpc and $\alpha\sim6 - 10$ at still larger radii. \cite{deason+13a}, proposed that the existence of the `break radius' in the Milky Way halo is associated with the `pile up' of stellar apocenters at a comparable Galactocentric distance. The observed existence of a break radius in the Milky Way halo and the absence of such a break in the Andromeda galaxy (M31) suggests that the latter had a more prolonged accretion history than the former.  More recently however, \cite{xue+15} and \cite{das+16a} showed with a sample of SEGUE K giants that if the flattening of the stellar-halo component is allowed to vary, a break is not required in the density profile.

The anisotropy profile of the stars preserves dynamical processes that affected the stellar orbits. If we consider the anisotropy parameter, $\beta=1-(\sigma_{\theta}^2 + \sigma_{\theta}^2)/2\sigma_r^2$, $\beta=-\infty$ for purely circular orbits and $\beta=1$ for purely radial orbits. \cite{chiba+98} obtained $\beta=0.52$ for a small sample of local halo red giants and RR Lyrae observed by the Hipparcos space mission. Using the SDSS Stripe 82 proper motions,  and  thus  going  deeper,  \cite{smith+09b} measured $\beta=0.69$  using $\sim 2000$  nearby subdwarfs. Combining  the  SDSS observations with the digitized photographic plate measurements, \cite{bond+10} increased the stellar  halo  sample  further  and derived $\beta=0.67$. For slightly larger volumes probed with more luminous tracers, similar values of $\beta\sim0.5$ were obtained \citep[see e.g.][]{deason+12a,kafle+12}. Beyond  15--20 kpc from the Sun, there have been several attempts  to  measure $\beta$ from just line-of-sight  velocity  measurements \citep[see e.g.][]{sirko+04,williams+15b}, although \cite{hattori+17} showed that this may depend heavily on the functional form assumed for the $\beta$ parameter. Only a few studies rely on proper motion measurements. These include a very small number of stars with proper motions measured with the Hubble Space Telescope finding a drop to $\beta\sim 0$ at $\sim20$ kpc \citep[see][]{deason+13b,cunningham+16}. \cite{das+16a} and \cite{das+16b} fit equilibrium dynamical models to K giants and blue horizontal branch (BHB) stars using proper motions from UCAC5. They find orbits becoming more isotropic going outwards in the K giant population, and possibly also in the BHB population. Both populations are found to be only mildly radially anisotropic in the inner halo.

The assembly history of the stellar halo is also reflected in its abundance structure. \cite{carollo+07} claimed a negative metallicity gradient, with the outer halo significantly more metal poor than the local halo. This was later confirmed by several studies \citep[e.g.][]{dejong+10,kafle+13,allende+14,chen+14}, with some also finding metal-poorer stars in retrograde motion and metal-richer stars in prograde motion \citep{carollo+07,kafle+13}. \cite{schonrich+14} argued however that metal-poorer stars can be seen at greater distances than metal-richer stars, and if this effect is not correctly included in the adopted selection function, a metallicity gradient can be erroneously inferred. The latest studies on the topic by \cite{xue+15} and \cite{das+16a}, who take the selection function into account in a sample of SEGUE K giant stars, only find a modest metallicity gradient at best.

The claims of a break in the density profile, a changing anisotropy profile, and a gradient in the metallicity have been interpreted as evidence for the existence of a dual halo, comprised of both {\it in-situ} and {\it accreted} components in the stellar halo. The metal-rich, in-situ stars dominate the inner halo, and metal-poor, accreted stars dominate the outer halo. A number of cosmological simulations support a similar composition of the stellar halo \citep[e.g.][]{zolotov+09,font+11,tissera+12,pillepich+15}. Although the stellar halo has generally been found to comprise old stars, $10$--$12\,$Gyr in age, born early in the Universe \citep[e.g.][]{jofre+11,kali+12}, \cite{nissen+10} and \cite{schuster+12} identified the presence of a high-$\alpha$ population of metal-poor stars in the Solar Neighbourhood with ages 2-3 Gyr larger than a low-$\alpha$ population at similar metallicities. The high-$\alpha$ halo stars were also found at smaller radii and heights and a range of eccentricities. The low-$\alpha$ halo stars, on the other hand, are clumped at eccentricities greater than 0.85. \cite{hawkins+14} further showed that a bifurcation appears in the age-metallicity relation such that in the low-metallicity regime the $\alpha$-rich and $\alpha$-poor populations are coeval, while in the high-metallicity regime the $\alpha$-rich population is older than the $\alpha$-poor population. This suggests that the $\alpha$-rich halo population, which has a shallow age-metallicity relation, could have formed in a rapid event with a high star formation rate (SFR) such as the thick disc of our Milky Way, while the $\alpha$-poor stars were formed in an environment with a slower chemical evolution timescale such as in dwarf spheroidal galaxies. Similar conclusions have been made by \cite{fernandez+18} and \cite{haywood+18} in their analysis of the chemical compositions and kinematics of the metal-rich nearby halo using data from the APOGEE survey combined with the second data release from Gaia \citep{gaia+18}. \cite{hayes+18} further showed that the metal-rich low-[Mg/Fe] halo stars were distinct in multiple abundance planes constructed with APOGEE DR14 data. 

A number of works published in the last year have associated the metal-rich, $\alpha$-poor inner halo specifically with the debris of a single, relatively massive stellar system that was accreted onto the Milky Way \citep{helmi+18,belokurov+18,mackereth+19}. \cite{helmi+18} examine stars from the cross-match between the second data release (DR2) from Gaia and the 14$^{\mathrm{th}}$ data release of APOGEE that are primarily counter-rotating (i.e. the $z$ component of angular momentum, $L_z$ is limited to $-1500<L_z<150$ kpc km/s) and loosely bound (i.e. $E>-1.8\times10^5$ km$^2$/s$^2$). They suggest that the selected stars belong to a single massive progenitor, `Gaia-Enceladus', due to the tightness of their sequence in the $\aFe$-$\FeH$ plane. The system has a large spread in metallicity, implying it did not form in a single burst but rather had an extended star formation history (SFH). Using chemical evolution models, they estimate star formation produced 0.3$\mathrm{M}_{\odot}$/year and lasted about two Gyr, implying that the progenitor has a stellar mass of $\sim6\times10^8 \mathrm{M}_{\odot}$, comparable to the Large and Small Magellanic Clouds. 

\cite{belokurov+18} used a large sample of Main Sequence stars with data from the first data release (DR1) from Gaia and SDSS to study the kinematic properties of the local stellar population within $\sim 10$ kpc from the Sun. They estimate distances from the photometry using the relations in \cite{ivezic+08} and use proper motions estimated from combining SDSS and Gaia DR1. They examine the tangential ($v_{\theta}$) and radial ($v_r$) components of the sample sliced by metallicity and vertical distance from the plane ($z$). They find two components in the metal-richer stars ($-1.33<\FeH<-1.00$), one co-rotating with the disk and with a small spread in $v_r$, and the other showing minimal rotation and a large spread in $v_r$. By examining a set of cosmological simulations, they claim that the very high radial anisotropy of the `Gaia Sausage' stars ($\beta\sim0.9$) is inconsistent with a series of minor accretion events but rather is the result of a massive satellite that sinks deeper into the potential well of the Milky Way as a result of dynamical friction. The radialization is then further enhanced in the presence of a growing disc as a result of loss in angular momentum. \cite{mackereth+19} came to similar conclusions based on their analysis of EAGLE simulations. They show that the median orbital eccentricities of debris are largely unchanged since merger time, implying that this accretion event likely happened at $z \sim 1.5$. Based  on Gaia DR2  data, \cite{myeong+18} found 12 halo GCs to be on highly eccentric orbits, at $e \sim 0.85$ thus making them consistent with an origin in a single massive accretion event. 

With the abundant spectroscopic and astrometric data for a large number of halo stars, we are now in a good position to estimate individual ages, and as such, obtain a complete picture of the nature of accreted halo stars. In this paper, we revisit the cross-match between Gaia DR2 and APOGEE DR14 and develop a new method for identifying a clean accreted sample of stars that is kinematically unbiased (within the selection function of the survey). We calculate new ages for these stars, and re-examine their distribution in a number of chemical abundance planes, age, and kinematics.

\section{The data}
Here we introduce the spectroscopic data that will be used to select accreted stars, and the kinematic data that will be used to further analyse the sample.

\subsection{Chemical abundances, spectral parameters, and line-of-sight velocities}
We use abundances and line-of-sight velocities measured by the near-infrared APOGEE survey. 
Data Release 14 \cite[DR14,][]{abol+18,holtzman+18} contains high S/N, moderate resolution ($R = \lambda/\Delta\lambda \sim22,500$)  spectra,  line-of-sight  velocities,  stellar  photospheric  parameters, and element abundances in up to 19 elemental species for over 270,000 stars in the H-band (1.5-1.7$\mu$m). Most stars are red giants, with a significant contribution from red dwarf stars. 

Spectra  are  reduced and analysed through the APOGEE data reduction pipeline \citep{nidever+15}, and  the  APOGEE  Stellar  Parameters  and  Chemical  Abundances Pipeline \citep[ASPCAP,][]{garcia+16}. ASPCAP uses a library of synthetic stellar spectra \citep{zamora+15} that is precomputed using a customised linelist \citep{shetrone+15} to measure stellar parameters, 19 element abundances and heliocentric line-of-sight velocities \citep{holtzman+18}. There is a two-step process. First,  the  stellar  parameters $\teff$, $\log g$, $v_{\mu}$ (micro-turbulent velocity), $\MH$, $\aM$, $\CFe$, and $\NFe$  are  determined  via  a  global  fit. The individual element abundances are then calculated by adjusting the $\MH$ ($\CFe$ and $\NFe$ for C and N, and $\aM$ for $\alpha$ elements) of the best-fit spectrum. The abundances  for species other than C and N can then be calibrated internally relative to open cluster observations, to account for systematic abundance variations with $\teff$. Surface gravities are calibrated using stars with independent asteroseismology determinations from the Kepler mission \citep{haas+10}. We apply the following quality control cuts, requiring:

\begin{enumerate}
\item{\texttt{ASPCAPFLAG = 0,}}
\item{\texttt{STARFLAG = 0,}}
\item{\texttt{M\_H, ALPHA\_M, MG\_FE, C\_FE, N\_H, MN\_FE, and AL\_FE must be known,}}
\item{\texttt{M\_H\_ERR, ALPHA\_M\_ERR, MG\_FE\_ERR, C\_FE\_ERR, N\_H\_ERR, MN\_FE\_ERR, and AL\_FE\_ERR must be less than 0.15,}}
\item{\texttt{TEFF > 4000 K,}}
\item{\texttt{LOGG > 0.5 dex,}}
\end{enumerate}
\noindent where \texttt{ASPCAPFLAG} and \texttt{STARFLAG} are flags in the APOGEE data that report potential issues with the star and/or with specific parameter determination process for that star. \texttt{M\_H, ALPHA\_M, MG\_FE, C\_FE, N\_H, MN\_FE}, and \texttt{AL\_FE} are the metallicity and abundance ratios, respectively,  as measured in the APOGEE pipeline. We require these particular abundance ratios to be known as they form the basis of our chemical selection criteria for studying the halo stellar populations.  We also impose a maximum uncertainty of 0.15 on these abundance ratios. Otherwise tails towards more extreme chemical abundances can appear, which can be difficult to capture in the Gaussian Mixture Models (GMM) discussed in Section \ref{sec:SelectingAccretedStars}. \texttt{TEFF} and \texttt{LOGG} are the APOGEE effective temperature and surface gravity, respectively. Spectral parameters and abundances are generally more reliable for the imposed ranges on effective temperature and $\log g$. Imposing these quality cuts reduces the sample to 136,212 stars. 

\subsection{Parallaxes, proper motions, and masses}
The  second  data release of the ESA-Gaia mission, Gaia DR2 \citep{gaia+18}, provides five-parameter astrometry (proper motions, positions and parallaxes) for over 1.3 billion objects in the Milky Way. Many improvements were made to the data-processing pipeline between Gaia DR1 \citep{lindegren+16} and DR2 that have reduced the uncertainty on astrometric parameters to $\sim$0.2 to 0.3 mas for stars in the middle of the covered range of magnitudes (going up to $\sim$2 mas for the faintest sources). We use the catalogue provided by \cite{sanders+18}, who perform a 5" radius cross-match between Gaia DR2 and six ground-based spectroscopic surveys (APOGEE, GALAH, Gaia-ESO, RAVE, LAMOST, SEGUE, altogether approximately three million stars) by utilising the Gaia proper motions and accounting for the respective epochs of the surveys. They extract the parallax, proper motions, and the uncertainty covariance matrix for the astrometry from the Gaia DR2 source catalogue. The authors supplement the catalogue with mass estimates for stars metal-richer than -1.5, based on an artificial neural network (ANN) trained on $\teff,\log g,\MH,\aM,\CFe, \mathrm{and}\,\NFe$. The relation between the mass of a giant star and $\CN$ has been discussed several times in the literature \citep[e.g.][]{masseron+15,martig+16,ness+16,das+19}. The minimum metallicity is set at -1.5 due to the lack in numbers of metal-poorer stars in the training set used to fit the neural network.


%
\begin{figure}
    \centering
    \includegraphics[scale=0.6]{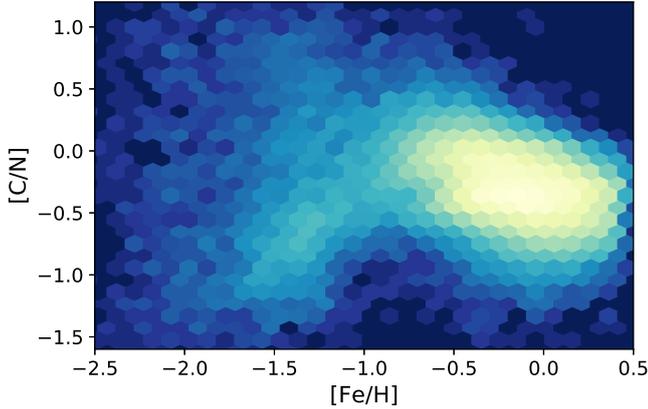}
    \caption{Two-dimensional distribution of APOGEE DR14 stars in the plane of $\FeH$ and $\CN$.}
    \label{fig:cnfeh}
\end{figure}

\section{Ages, distances, and orbital parameters}
\cite{sanders+18} employed the Bayesian isochrone pipeline of \cite{burnett+10} to derive the distance, metallicity, mass, age, and line-of-sight extinction of each star from the combined catalogues and the neural network mass estimates for stars metal-richer than -1.5. However, recent work \citep{lagarde+18,shetrone+19} suggests that efficient thermohaline mixing in metal-poor stars may lead to a large spread in $\CN$ at low metallicities. At these low metallicities, inhomogeneities in the interstellar medium (ISM) \citep[e.g.][]{revaz+12} also become increasingly important, e.g. stars formed close to SNII will have very different abundances from those further away. Figure \ref{fig:cnfeh} shows the distribution of the APOGEE DR14 stars in the $\FeH$-$\CN$ plane. $\CN$ decreases with metallicity along two sequences for stars metal-richer than approximately -0.8, likely the thick and thin disks of the Milky Way. As metallicity is roughly correlated with age, this implies that $\CN$ is a good age indicator in red giants metal-richer than -0.8\footnote{The correlation arises because $\CN$ is a good estimator of mass, and in red giants, mass is a measure of the time to complete the Main Sequence. The Main Sequence is where they spend most of their life.}. Metal-poorer stars are likely to mainly belong to the halo and therefore originate in different stellar systems. We therefore do not expect a regular sequence with metallicity as seen in the thick and thin disk stars. However the scatter in their $\CN$ is far greater than seen in the thick and thin disk stars, which is unexpected for stars that should generally be old. We therefore recalculate ages, distances, and orbital parameters for APOGEE DR14 stars metal-poorer than -0.8 and metal-richer than -1.5 using a method described below.

\subsection{The Bayesian isochrone pipeline}
We use the Bayesian isochrone pipeline presented in MADE \citep{das+19}. The parameters, $\phi^i$, of the star $i$ estimated from the Bayesian isochrone pipeline are given by
\begin{equation}
\phi^i = (\mathcal{M}^i,\tau^i,\MH^i,s^i),
\end{equation}
where $\mathcal{M}$ is the {\it initial} mass, $\tau$ is the age, $\MH$ is the metallicity, and $s$ is the distance. These parameters predict the following observed properties of the star $i$
\begin{equation}
v^i = (m^i, H^i, (J-K_s)^i,\varpi^i),
\end{equation}
where $m$ is the current mass, $H$ is the $H$-band magnitude, $J-K_s$ is a colour, and $\varpi$ is the parallax. The accompanying measured observed stellar properties are denoted by  $\tilde{v}^i$. These are associated with measurement uncertainties, $\tilde{\rho}^i$.
\begin{table}
 \centering
  \caption{Parameters of the four-component Milky Way prior. \label{tab:priorpars}}
  \begin{tabular}{lll}
  	\hline
	Component		&Parameter 			   			&Value \\	
	\hline	
	Bulge			&$\mu_{\MH,1}$/dex	   			&-0.3\\
					&$\sigma_{\MH,1}$/dex   		&0.3\\
					&$\mu_{\tau,1}$/Gyr	   			&5.0\\
					&$\sigma_{\tau,1}$/Gyr  		&5.0\\		
					&$q$							&0.5\\
					&$\gamma$					    &0.0\\
					&$\delta$						&1.8\\
					&$r_0$/kpc						&0.075\\
					&$r_t$/kpc						&2.1\\
	Thin disc		&$\mu_{\MH,2}$/dex	   			&0.0\\
					&$\sigma_{\MH,2}$	   			&0.2\\		
					&$R_{\mathrm{d},2}/$kpc 		&2.6\\
					&$z_{\mathrm{d},2}/$kpc 		&0.3\\
	Thick disc		&$\mu_{\MH,3}$/dex	   			&-0.6\\
					&$\sigma_{\MH,3}$/dex   		&0.5\\
                    &$\mu_{\tau,3}$/Gyr	   			&10.\\
					&$\sigma_{\tau,3}$/Gyr   		&2.\\
					&$R_{\mathrm{d},3}/$kpc 		&3.6\\
					&$z_{\mathrm{d},3}/$kpc 		&0.9\\
	Stellar halo	&$\mu_{\MH,4}$/dex	   			&-1.6\\
					&$\sigma_{\MH,4}$/dex   		&0.5\\
					&$\mu_{\tau,4}$/Gyr	   			&11.0\\
					&$\sigma_{\tau,4}$/Gyr  		&2.0\\	
	\hline	
	\end{tabular}
\end{table}

Applying Bayes' law to each star gives
\begin{equation}\label{eqref:bayes_modpars_individstar}
p(\phi^i|\tilde{v}^i,l^i,b^i) =\frac{p(\tilde{v}^i|\phi^i,l^i,b^i) p(\phi^i|l^i,b^i)}{p(\tilde{v}^i)},
\end{equation} 
where $(l^i,b^i)$ are the predicted sky positions of star $i$ in Galactic coordinates (which we assume to be the same as the observed sky positions), and $p(\tilde{v}^i)$ is an unimportant normalization. The apparent magnitudes of star $i$ are extinction-corrected using the state-of-the-art \texttt{Combined15} map compiled by \cite{bovy+16b} in the \texttt{mwdust} package. The likelihood of the star's observed properties, $p(\tilde{v}^i|\phi^i,l^i,b^i)$, is assumed to be the product of the separate likelihoods. Each likelihood is represented by a Gaussian distribution
\begin{equation}
G(\tilde{v}_j^i,v_j^i,\tilde{\rho}_j^i) = \frac{1}{\sqrt{2\pi}\tilde{\rho}_j^i}\exp\left(-\frac{(\tilde{v}_j^i-v_j^i)}{2(\tilde{\rho}_j^i)^2}^2\right).
\end{equation}
Thus
\begin{equation}
p(\tilde{v}^i|\phi^i,l^i,b^i) = \prod_j G(\tilde{v}_j^i,v_j^i,\tilde{\rho}_j^i).
\end{equation}
The pipeline employs PARSEC isochrones v1.1 \citep[assuming a mass-loss efficiency, $\eta=0.2$, ][]{bressan+12} evaluated for 57 metallicities ranging between -2.192 and 0.696, and 353 ages ranging between $\log_{10}\tau = 6.60$ and 10.12 (i.e. a spacing of $\Delta \log_{10}\tau=0.01$) for which we create a dictionary of interpolants in {\sc Python} that returns luminosity, $\log g$, $T_{\mathrm{eff}}$ and apparent magnitudes, given the metallicity, age, and mass of a star. We use a prior $(\phi^i|l^i,b^i)$ informed by the Milky Way model presented in \cite{das+19} (the superscript $i$ is omitted in the following)
\begin{equation}
\begin{split}
p(\mathcal{M},\tau,\MH,s|l,b) = s^2\epsilon(\mathcal{M})\sum_{k=1}^{4}p_k(\MH)p_k(\tau)p_k(R,z),
\end{split}
\end{equation}
where $k=1,2,3,4$ correspond to a bulge, thin disc, thick disc, and stellar halo respectively. The $s^2$ term accounts for the Jacobian of the transformation of spatial coordinates, and $\epsilon(\mathcal{M})$ is the Kroupa \citep{kroupa+93} initial mass function (IMF)
\begin{equation}
	\epsilon(\mathcal{M}) = 
	\begin{cases}
		0.035\mathcal{M}^{-1.5} &\mathrm{if} \, 0.08 \leq \mathcal{M/M_{\odot}} < 0.5,\\
		0.019\mathcal{M}^{-2.2} &\mathrm{if} \, 0.5 \leq \mathcal{M/M_{\odot}} < 1.0,\\
		0.019\mathcal{M}^{-2.7} &\mathrm{if} \, \mathcal{M/M_{\odot}} \geq 1.0\,.
	\end{cases}
\end{equation}
The priors for the four components are\\

\noindent Bulge ($k=1$):
\begin{equation}
\begin{split}
p_1(\MH) &= G(\MH,\mu_{\MH,1},\sigma_{\MH,1}),\\ 
p_1(\tau) &= G(\tau,\mu_{\tau,1},\sigma_{\tau,1}),\\
p_1(R,z) &\propto \frac{(1+m)^{(\gamma-\delta)}}{m^{\gamma}}\exp[-(mr_0/r_t)^2],\\
\mathrm{where}\,m(R,z) &= \sqrt{(R/r_0)^2 + (z/qr_0)^2}.
\end{split}
\end{equation}
\noindent Thin disc ($k=2$):
\begin{equation}
\begin{split}
p_2(\MH) &= G(\MH,\mu_{\MH,2},\sigma_{\MH,2}),\\ 
p_2(\tau) &\propto
\begin{cases}
	\exp(\frac{\tau}{8.4}) 
    &\mathrm{if} \, \tau/\mathrm{Gyr} \le 8 \\
    2.6\exp\left(-0.5\frac{(\tau-8)^2}{1.5^2}\right) 
    &\mathrm{if} \, \tau/\mathrm{Gyr} > 8 
\end{cases}\\
p_2(R,z) &\propto \exp\left(-\frac{R}{R_{\mathrm{d},2}}-\frac{|z|}{z_{\mathrm{d},2}}\right).
\end{split}
\end{equation}

\noindent Thick disc ($k=3$):
\begin{equation}
\begin{split}
p_3(\MH) &= G(\MH,\mu_{\MH,3},\sigma_{\MH,3}),\\
p_3(\tau) &= G(\tau,\mu_{\tau,3},\sigma_{\tau,3}),\\ 
p_3(R,z) &\propto \exp\left(-\frac{R}{R_{\mathrm{d},3}}-\frac{|z|}{z_{\mathrm{d},3}}\right).
\end{split}
\end{equation}
\noindent Stellar halo ($k=4$):
\begin{equation}
\begin{split}
p_4(\MH) &= G(\MH,\mu_{\MH,4},\sigma_{\MH,4}),\\ 
p_4(\tau) &\propto \, G(\tau,\mu_{\tau,4},\sigma_{\tau,4}) \\\
p_4(R,z) &\propto r^{-3.39}.
\end{split}
\end{equation}
The parameters of the prior are given in Table \ref{tab:priorpars}. The thin disc is normalized to have a local density of 0.04 $M_{\odot}$pc$^{-3}$ \citep{bovy17}. The thick disc and stellar halo are normalized so that their local densities have ratios of 0.04 and 0.005 with the thin disc, respectively \citep{bland+16,bovy17}. The bulge component is normalized to have a central density of 35.45 $M_{\odot}$pc$^{-3}$ \citep{robin+12}. An overarching prior is imposed that constrains metallicities and ages to the range covered by the isochrones.

We calculate $p(\phi^i|u^i)$ on a grid of all 353 ages, $\tau^i$,  all metallicities, $\MH^i$, lying within $3\sigma$ of the measured metallicity, 2000 initial masses, $\mathcal{M}^i$, ranging between the minimum and maximum mass of the relevant isochrone, and 30 distances, $s^i$, based on a linear grid of parallaxes lying within 3$\sigma$ of the the measured parallax. We calculate the first and second moments of the logarithm of age, $\log_{10}\tau$, metallicity, [M/H], logarithm of the current mass, $\log_{10}m$, and the distance modulus, $\mu$ as our outputs and output uncertainties, e.g.
\begin{equation}
\begin{split}
\langle \log_{10}\tau \rangle = \Big(\int \,\mathrm{d}\phi\,\log_{10}\tau \,p(u|\phi,l,b) p(\phi|l,b)\Big) \Big/\\ \Big(\int \mathrm{d}\phi\,p(u|\phi,l,b)p(\phi|l,b)\Big).
\label{eq:moments}
\end{split}
\end{equation}

The primary difference between the Bayesian isochrone pipeline presented here and that of \cite{sanders+18} is that we assume a point estimate for the line-of-sight extinction from previous studies rather than make a new estimate. In applying the pipeline to stars with metallicities between -1.5 and -0.8, we also account for the systematic parallax offset of 0.03 mas \citep{lindegren+18}, because this may have a minor effect on the ages of halo stars for which parallaxes are smaller. The effect of the offset is small and we therefore do not worry about stars metal-poorer than -1.5, for which the offset has not been accounted for.
\begin{figure}
    \centering
    \includegraphics[scale=0.6]{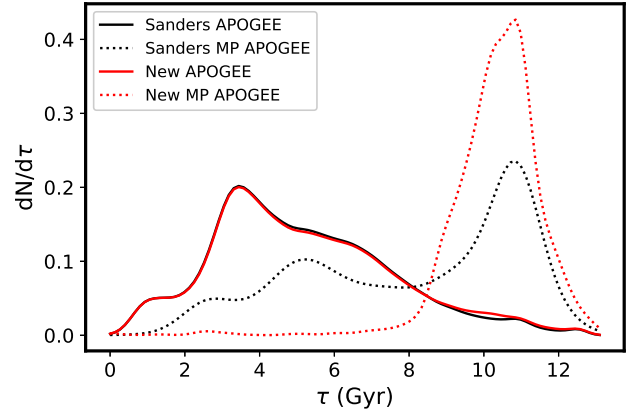}
    \caption{One-dimensional age distributions for all APOGEE stars (solid), and stars metal-poorer than -0.8 (dotted) in \protect\cite{sanders+18} (black) and the present work (red). The distributions have been normalized so that the area under each is one.}
    \label{fig:ages}
\end{figure}

The isochrone pipeline successfully determines outputs for $203,127$ stars in the overlap between APOGEE DR14 and Gaia DR2. Figure \ref{fig:ages} compares the one-dimensional age distributions determined here and in \cite{sanders+18} for these stars. We further compare the age distribution of the stars that are metal-poorer than -0.8. The distributions  are very similar when we look at the whole sample because it is dominated by young, metal-rich disc stars. But when looking at just the metal-poor stars, which pick out mainly thick-disc and halo stars, we find that \cite{sanders+18} predict a significantly higher number of young stars.
\begin{figure*}
    \centering
    \includegraphics[scale=0.7]{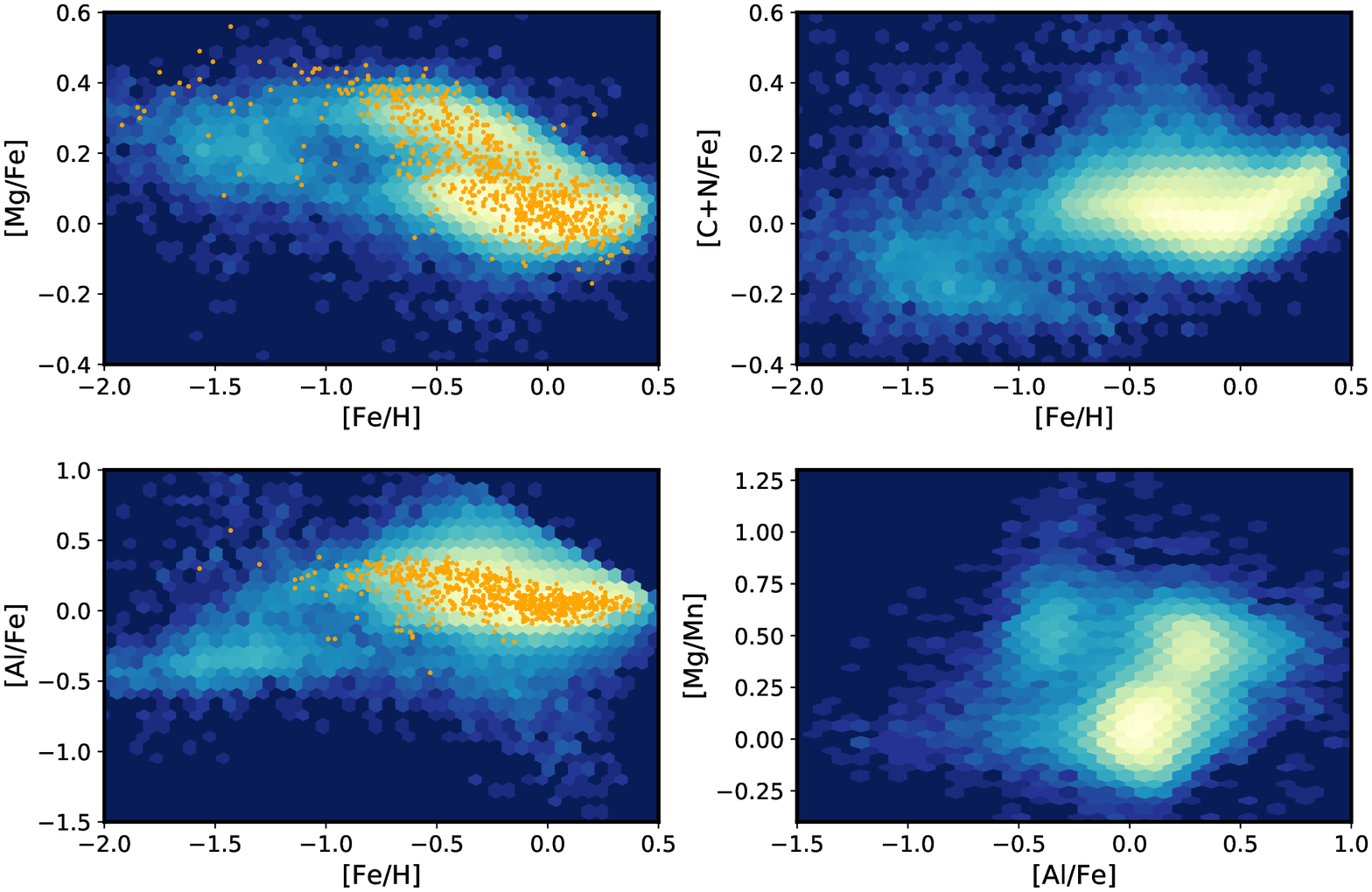}
    \caption{Two-dimensional abundance distribution of APOGEE DR14 stars in the plane of $\FeH$ and $\MgFe$ (top left), $\FeH$ and $\AlFe$ (bottom left), $\FeH$ and $\CplusNFe$ (top right), and $\CN$ and $\FeH$ (bottom right). Filled orange circles show abundances for 714 kinematically-selected disc F and G dwarfs and subgiants from \protect\cite{bensby+14}.}
    \label{fig:abundances}
\end{figure*}

\subsection{Estimating orbital parameters}
We make estimates of orbital parameters, called actions, using the AGAMA software package of \cite{vasiliev+19} from six-dimensional phase-space coordinates in the APOGEE DR14-Gaia DR2 cross-match, and assuming the gravitational potential of \cite{piffl+14}. In a near-integrable potential, actions, which are constants of motion, efficiently package the orbital properties of each of the stars into three labels. Assuming an axisymmetric potential, the three actions are the radial action $J_r$, vertical action $J_z$, and $z$ component of angular momentum, $L_z$. The radial and vertical actions approximately describe excursions of an orbit in the radial and vertical directions. We apply a last quality control cut to create our final catalogue by requiring:
\begin{enumerate}
\item{\texttt{Age, $J_r$, $J_z$, and $L_z$ are known,}}
\end{enumerate}
Our final sample contains 132,380 stars. Typical uncertainties are 0.03 dex in $\MH$, 0.02 in $\MgFe$, 0.06 in $\AlFe$, 0.04 in $\CFe$, and 0.05 in $\NFe$. Typical uncertainties in mass are 5\% and in age are 18\%.

\section{Selecting accreted stars in the Milky Way}
\label{sec:SelectingAccretedStars}
Here we examine the Milky Way stars in a number of abundance planes, and use a Gaussian Mixture Model to identify stars that are likely to be accreted.

\subsection{Abundance planes}
\cite{hawkins+15} present a chemical abundance distribution study in 14 $\alpha$, odd-$Z$, even-$Z$, light, and Fe-peak elements of approximately $3200$ intermediate-metallicity giant stars from APOGEE-DR12. They suggest a set of chemical abundance planes constructed from combinations of $\aFe$, $\CplusNFe$, $\AlFe$, and $\MgMn$ that may be able to chemically label the Galactic components in a clean way independent of kinematics.  

Figure \ref{fig:abundances} presents some of the abundance planes considering these abundance ratios. We overplot abundances for 714 F and G dwarf and subgiant stars in the Solar Neighbourhood from \cite{bensby+14}, who conducted a high-resolution ($R = 40,000 - 110,000$) spectroscopic study using spectra from a number of telescopes. The sample was kinematically selected to trace the Galactic thin and thick disks to their extremes, and therefore should reflect the underlying range of abundances expected for the Milky Way thin and thick discs. Typical uncertainties are likely similar between the two samples, but are difficult to quantitatively compare in detail \citep[see, e.g, ][for a recent review]{jofre+18} because the systematic uncertainties are likely to differ. This is a consequence of firstly the abundances in the sample of \cite{bensby+14} being extracted from lines in the optical wavelength range while APOGEE abundances are from lines in the infrared wavelength range. We expect the resulting abundance trends in the discs from both samples to be similar however.

\begin{figure*}
    \centering
    \includegraphics[scale=0.7]{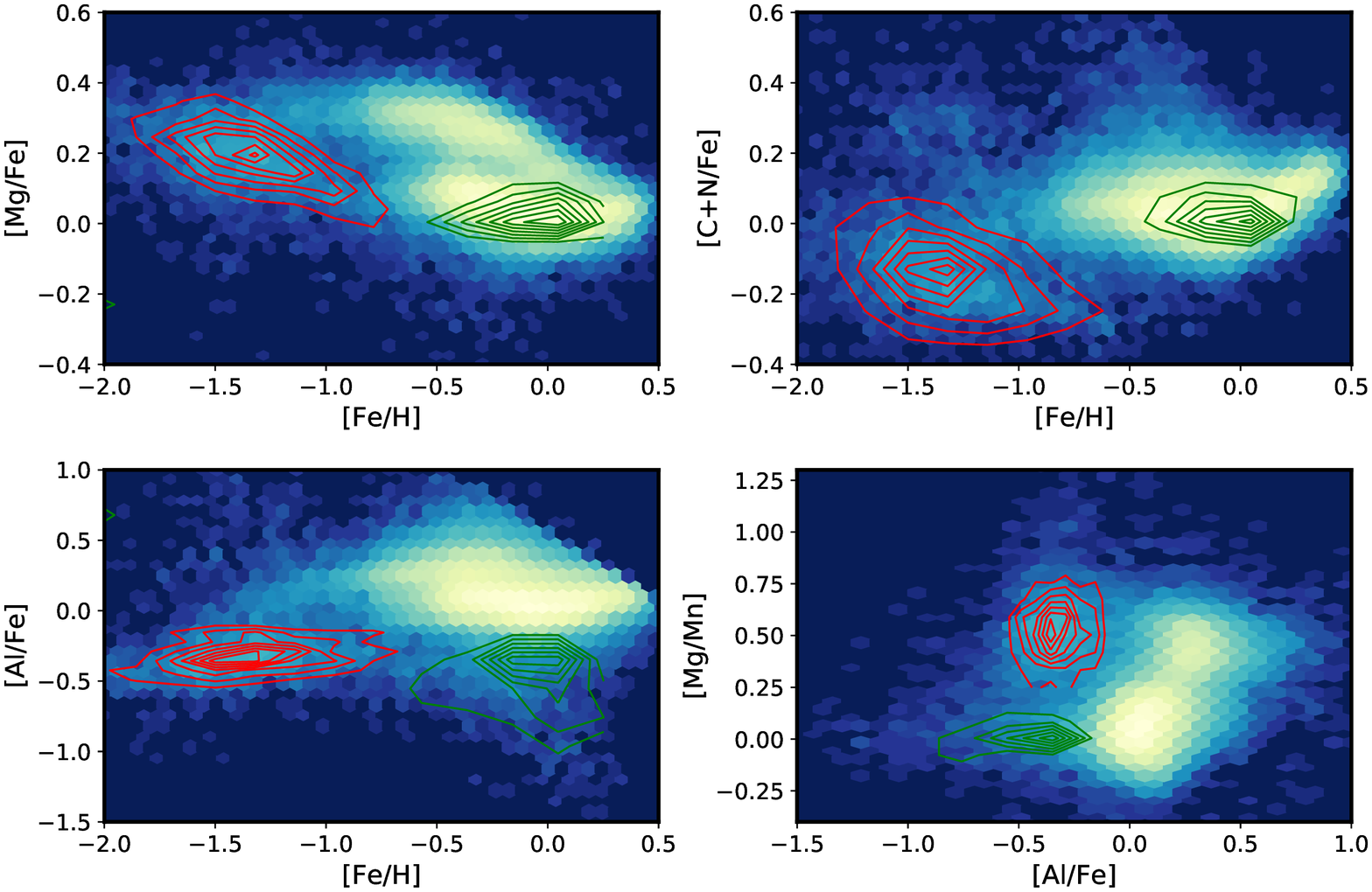}
    \caption{Two-dimensional abundance distribution of APOGEE DR14 stars in the plane of $\FeH$ and $\MgFe$ (top left), $\FeH$ and $\AlFe$ (bottom left), $\FeH$ and $\CplusNFe$ (top right), and $\CN$ and $\FeH$ (bottom right). The contours show the distributions of stars belonging to the two low-$\AlFe$ Gaussian components found by the GMM. The red contours represent the `blob', and the green contours, likely low $\AlFe$ thin-disc stars.}
    \label{fig:abundances_GMM}
\end{figure*}

The $\FeH$-$\MgFe$ plane shows two overdensities at around solar metallicity. The locations of the disk stars selected by \cite{bensby+14} imply the two overdensities are likely to be the thin ($\alpha$-poor) and thick ($\alpha$-rich) discs of the Milky Way. These overdensities are well known and have already been characterized in detail \citep[for example in][]{hayden+14}. There is also a faint overdensity at intermediate $\MgFe$ and lower $\FeH$. These stars have a slightly lower $\alpha$ abundance than the thick-disc stars and are at significantly lower metallicities. Discussions about this metal-poor and relatively $\alpha$-poor population has notably been motivated by \cite{nissen+10}, who conjectured that these stars may have originated in external galaxies that were accreted on to our Milky Way. Some evidence for this lies in the slopes of the sequences traced by each of the overdensities. $\alpha$ elements are produced in the cores of short-lived massive stars during the post-Main-Sequence burning phase. They are dispersed in the interstellar medium (ISM) via Type II, core-collapse supernovae (SNII). Some iron is also produced. Type 1a supernovae (SN1a) of longer-lived stars are the main production site for iron-peak elements, but only contribute a minimal amount of $\alpha$ elements. Therefore as a stellar population evolves, the $\alpha/$Fe decreases with $\FeH$, as can be seen in both the thick and thin discs. For massive systems, enrichment by iron happens on a shorter timescale compared to smaller systems due to a higher SFR and a higher ability to retain the expelled material in the potential well. Therefore the slope of the sequence of stars in [$\alpha$/Fe]-$\FeH$ is shallower in the metal-poorer overdensity. Assuming that the IMF is universal, the oldest stars in each system will start as very metal-poor and high in $\alpha$. This can be clearly seen in the sample from \cite{bensby+14} where the most $\alpha$-rich stars extend down to very low metallicities.

\begin{figure*}
    \centering
    \includegraphics[scale=0.6]{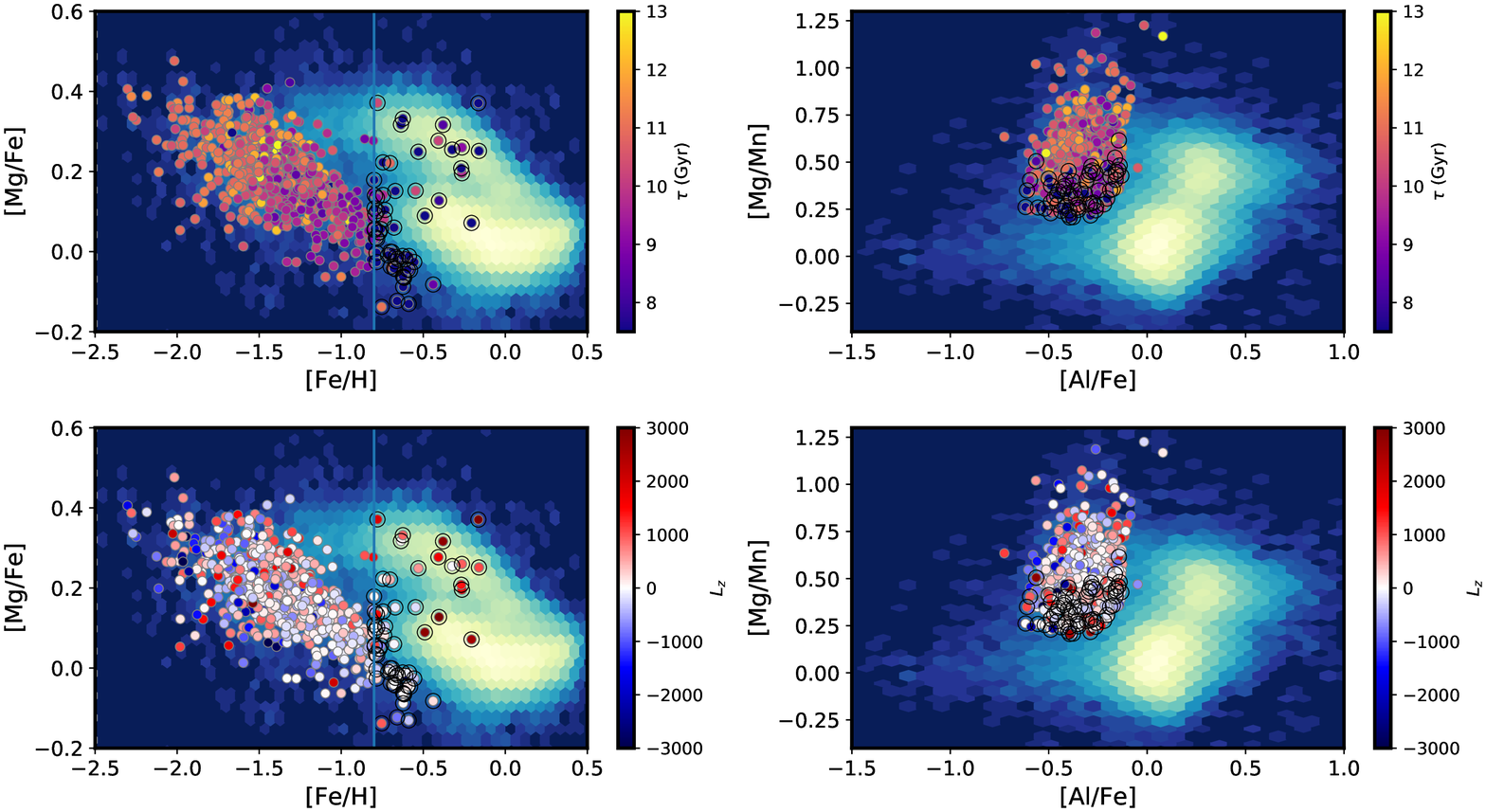}
    \caption{Density maps of APOGEE DR14 stars in the $\MgFe$-$\FeH$ (left) and $\AlFe$-$\FeH$ (right), planes with the ages (top) and $z$ component of angular momentum (bottom) of the blob stars superimposed. The vertical blue line shows the additional criterion imposed to remove metal-rich contaminants in the blob. The black circles highlight the blob stars that are removed as a result.}
    \label{fig:abundances_ages_actions}
\end{figure*}

The $\FeH$-$\CplusNFe$ plane shows a single overdensity around zero metallicity, and a second overdensity at lower metallicity and lower $\CplusNFe$. APOGEE stars are giants, and therefore their surface C and N abundances have been partially affected during their evolution as a result of dredge-up processes, as discussed earlier. However, the initial $\CplusNFe$ is approximately conserved throughout the evolution of those stars \citep[see ][]{masseron+15}, due to only a slight change in $^{16}$O. Therefore, $\CplusNFe$ should depend on metallicity and environment. C is mostly made by He burning in the cores of stars and dispersed in the ISM via SNII at very low metallicity as well as asymptotic giant branch (AGB) stars at around $\sim$-1.50 dex in metallicity. Hence, $\CplusNFe$ is expected to star decreasing as SN1a start dominating, which do not produce C. It may then start increasing as a result of production in AGB stars. There trends are not clear in the overdensities, but the metal-poorer overdensity appears to have lower $\CplusNFe$ at similar metallicities, which is expected for stars born in less massive systems.

The $\FeH$-$\AlFe$ plane is very similar to the $\FeH$-$\CplusNFe$ plane. The locations of the disc stars selected by \cite{bensby+14} in this plane show that the metal-richer overdensity is likely to be the discs of the Milky Way. $\AlFe$ decreases with metallicity in the sample from \cite{bensby+14}, but it is unclear what is happening at lower metallicities than -1.0. Na and Al are thought to be produced in the cores of stars and dispersed in the ISM via SNII. However, according to \cite{kobayashi+06}, the production quantities of those elements is strongly dependent on the initial C and N in the gas cloud that forms the stars. Therefore, it is expected that Na and Al are primarily correlated with C+N as is observed in the case of Al. Na and Al are also expected to be partially produced by AGB stars at intermediate metallicities \citep{nomoto+13}. Moreover, since SNIa do not produce Al and Na as efficiently as Fe, the $\AlFe$ and [Na/Fe] tend to decrease towards higher metallicities, which can be clearly seen in the sample from \cite{bensby+14}. \cite{nissen+10} demonstrate the effective ability of Na and Al to characterize accreted stars born in less massive systems, where, as with C+N, their overall abundances are lower. At these metallicities, typical APOGEE [Na/Fe] abundance uncertainties in this metallicity range approach $\sim \pm$0.20 dex. In contrast, the precision of Al abundances in the APOGEE data is effectively very high, $\sim \pm0.06$, and thus offer an alternative to Na. 

In summary, the abundance ratio of the $\alpha$ elements, generally decrease with metallicity, but at rates that depend on the SFR in the local environment. The overall abundances of C+N and Al of stars depend on the mass of the system in which they were born. Therefore at lower metallicities the $\aFe$-$\FeH$ sequences of stars born in different environments may overlap, but they diverge towards higher metallicities. Also low-$\AlFe$ stars are expected in both accreted systems, and in particular thin-disc, metal-rich stars, but their $\alpha$ abundances should be different. Therefore, we examine APOGEE DR14 stars in the $\MgMn$-$\AlFe$ plane (Figure \ref{fig:abundances}). We choose $\MgMn$ rather than $\MgFe$ as Mn is a pristine tracer of SN1a, unlike iron, which has other formation channels in stellar evolution. This plane appears to clearly separate out high-$\alpha$ thick-disc stars from high-$\alpha$ accreted stars \citep[][see further discussion in]{hawkins+15}, due to the difference in $\AlFe$. It also appears to separate out low-$\alpha$ thin-disc stars from low-$\alpha$ accreted stars due to their difference in $\MgMn$. 

\subsection{Applying a GMM}
We formalize the identification of the overdensities in the $\MgMn$-$\AlFe$ plane using the GMM-fitting routines available in {\sc Python}'s \texttt{skikit-learn} package. A GMM is a probabilistic model that assumes all the data points are generated from a mixture of a finite number of Gaussians with unknown parameters. We explore GMMs with up to 20 components, and find that the Bayesian Information Criterion (BIC) favours a model with 14 components. A star is assigned to a component if the probability of belonging to it is greater than 0.7. There is a cluster of components centred on what is likely to be the thick disc (i.e. they select high-$\alpha$, high-$\AlFe$ stars, and a second cluster of components centred on what is likely to be the thin disc (i.e. they select low-$\alpha$, high-$\AlFe$ stars). Two components are centred on lower $\AlFe$, one at lower values of $\MgMn$, and one at higher values. Figure \ref{fig:abundances_GMM} replots the density maps of Figure \ref{fig:abundances} with contours of stars belonging to the two low-$\AlFe$ Gaussian components overplotted.  As explained above, stars belonging to the component with lower values of $\MgMn$ are likely to be the low-$\AlFe$ thin-disc stars, also identified in \cite{hawkins+15}  through their kinematics. Like other thin-disc stars, these stars are low in $\MgMn$ because they are born after the ISM has been enriched by Mn 
expelled by Type 1a SNe. They are low in $\AlFe$, and therefore likely to comprise a mixture of accreted stars and metal-richer thin-disc stars.

\begin{figure}
    \centering
    \includegraphics[scale=0.6]{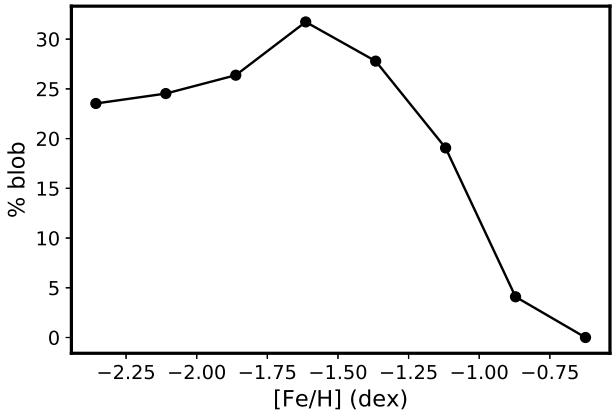}
    \caption{The variation in the fraction of APOGEE DR14 stars to the blob with metallicity.}
    \label{fig:blobinapogee}
\end{figure}

The high-$\MgMn$, low-$\AlFe$ component or `blob' is likely to be a relatively pure sample of accreted stars. We examine it in more detail in Section \ref{sec:blob}. There are 856 stars allocated to the blob.

\begin{figure}
    \centering
    \includegraphics[scale=0.6]{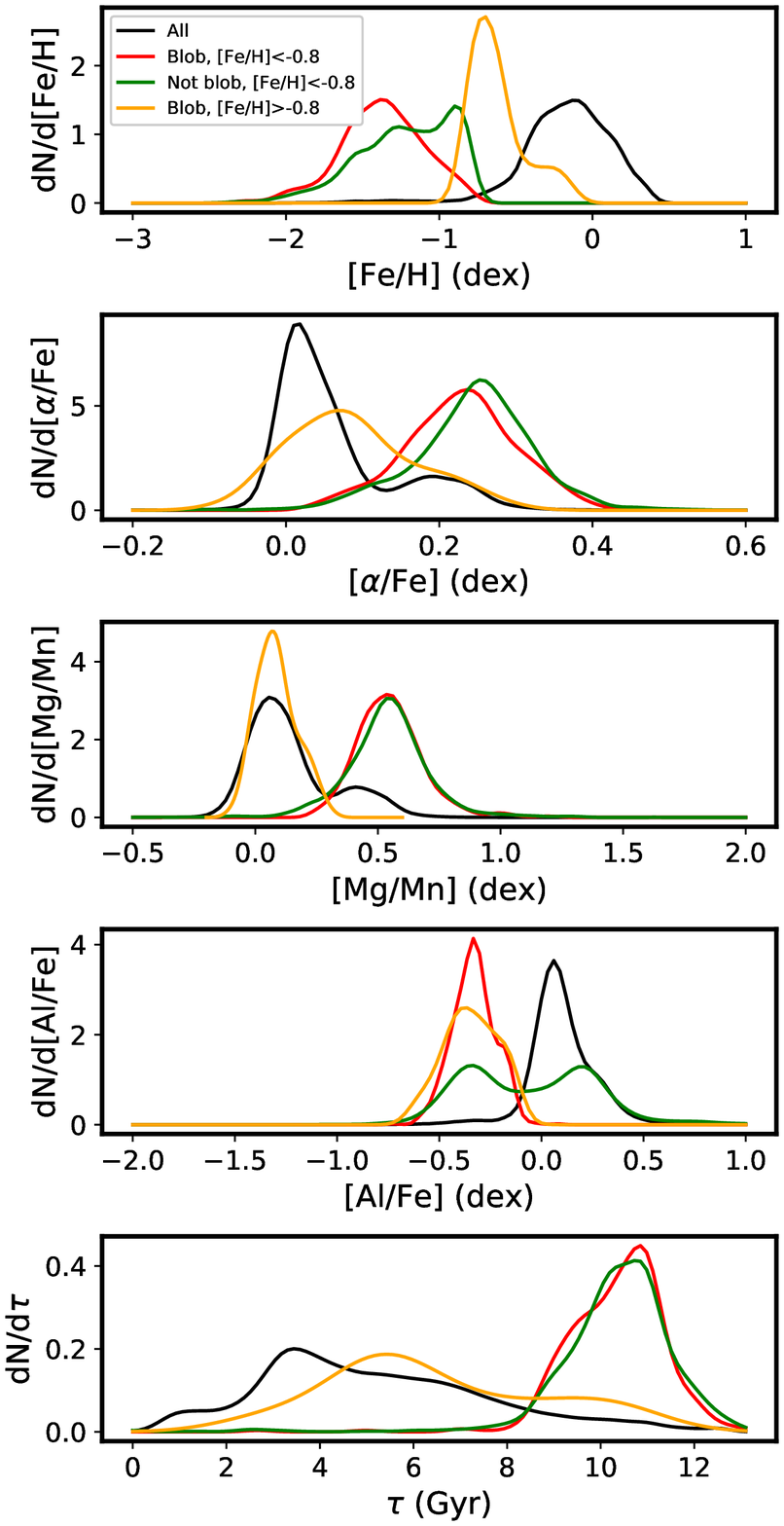}
    \caption{Going down from the top: kernel density estimates of the 1-D distributions of metallicity, $\alpha$-abundance, $\MgMn$, $\AlFe$, and age for all APOGEE DR14 stars (black, solid line), the blob stars without the metal-rich contaminants (red, solid line), metal-poor stars not belonging to the blob (green), and the metal-rich contaminants allocated to the blob (orange). The distributions are normalized so that the area under each is one.}
    \label{fig:blob1ddist_ageabundance}
\end{figure}

\begin{figure*}
    \centering
    \includegraphics[scale=0.6]{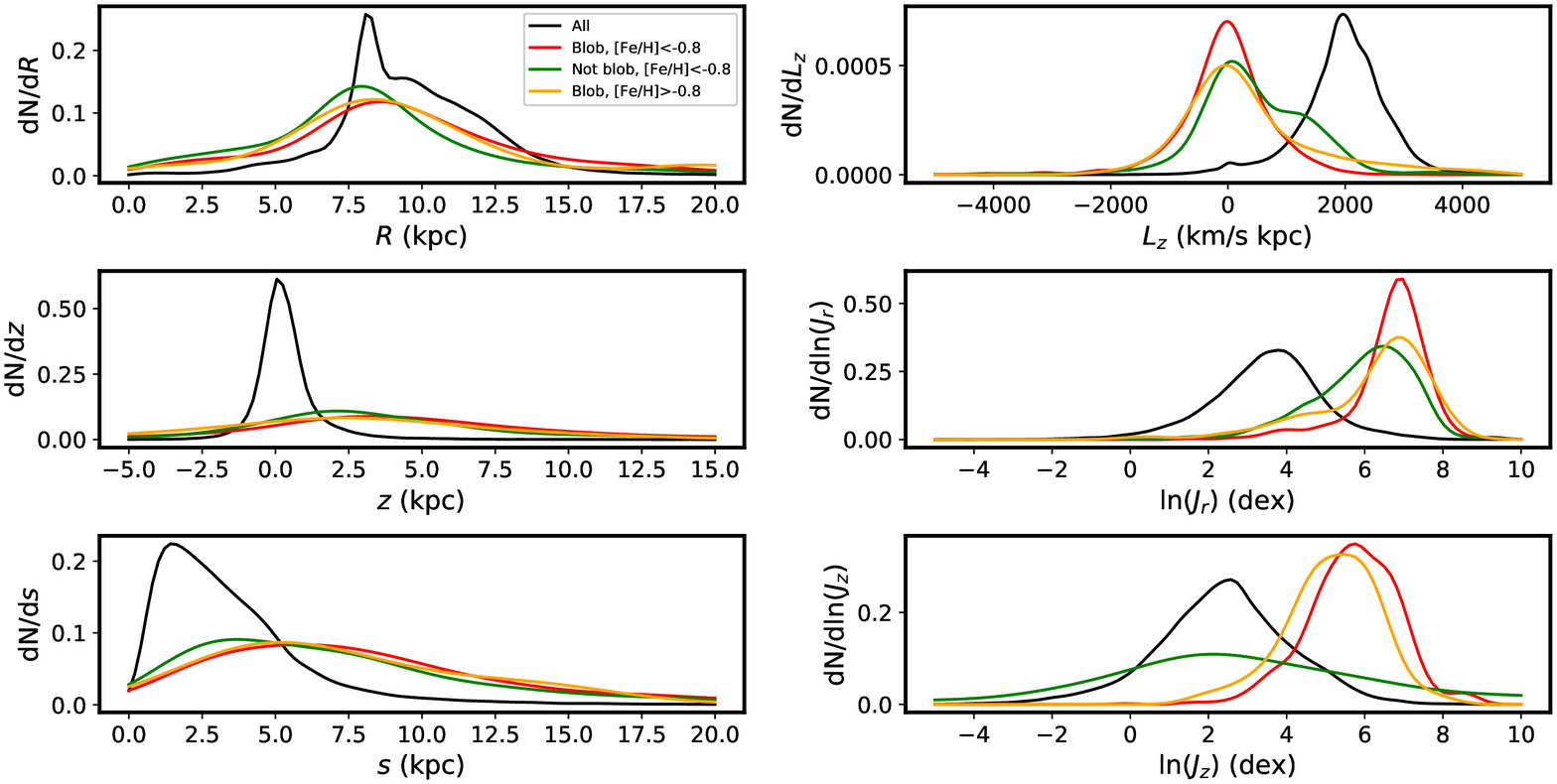}
    \caption{Going anti-clockwise from top left: kernel density estimates of the 1-D distributions of $R$, $z$, $s$, $\ln J_z$, $\ln J_R$, and the $z$ component of angular momentum for all APOGEE DR14 stars (black, solid line), the blob stars without the metal-rich contaminants (red, solid line), metal-poor stars not belonging to the blob (green), and the metal-rich contaminants allocated to the blob (orange). The distributions are normalized so that the area under each is one.}
    \label{fig:blob1ddist_kinematics}
\end{figure*}

\section{The blob stars}\label{sec:blob}
Here we examine the properties of the stars belonging to the blob\footnote{A catalogue of the spectroscopic, age, and kinematic properties of the blob stars can be downloaded from \url{https://drive.google.com/open?id=1TYncwKeWlBx7F2b8Y8KN2LWyMKpSKKIk.}}.

\subsection{Ages and kinematics in the abundance planes}
Figure \ref{fig:abundances_ages_actions} shows the location of the blob stars in the $\MgFe$-$\FeH$ (left) and $\MgMn$-$\AlFe$ (right) abundance planes, coloured by age (top) and the $z$ component of angular momentum. It is obvious from the $\MgFe$-$\FeH$ plots that some stars that were allocated to the blob are disc contaminants, either because they are younger than $\sim$7 Gyr and/or because they are co-rotating with the disc. They are low in $\AlFe$ and $\MgMn$ and are therefore allocated to the blob, but are comparatively metal-rich. They can be mainly removed by imposing an additional criterion in metallicity, i.e. $\FeH<-0.8$. 61 of the 856 stars allocated to the blob are suspected contaminants (i.e. $\sim 7\%$). We continue with this modified definition of the blob stars, and revisit the properties of the metal-rich contaminants in the blob in Section \ref{ssec:1DdistBlob}.

The remaining blob stars follow a sequence in the $\MgFe$-$\FeH$ plane, i.e. metal-richer stars are generally poorer in $\alpha$ elements, similar to that seen in the thick and thin discs, but over a larger range of metallicities and $\MgFe$. $\AlFe$ does not change as significantly over the same range in metallicities. The blob stars with the highest $\alpha$ and lowest metallicity are around 13 Gyr, and metal-richer stars are as young as 7 to 8 Gyr. This correlation is not in the Milky Way model prior that was used to calculate the new ages, and therefore must be driven by the data. The bottom panel of plots shows the $z$ component of angular momentum, $L_z$, of the blob stars. Both co-rotating and counter-rotating blob stars are found throughout the sequence, however the metal-richer stars may have a higher proportion of counter-rotating stars.

\subsection{Contribution to the metal-poor population of APOGEE}
Figure \ref{fig:blobinapogee} shows how the proportion of APOGEE DR14 stars belonging to the blob depends on metallicity. The contribution increases from a metallicity of around -2.5 at $\sim24\%$ to $\sim32\%$ at a metallicity just below -1.5. It then decreases towards solar metallicity. Therefore a significant fraction of the metal-poor stars in APOGEE DR14 belong to the blob.

\subsection{1-D abundance, age, and kinematic distributions}
\label{ssec:1DdistBlob}
Figures \ref{fig:blob1ddist_ageabundance} and \ref{fig:blob1ddist_kinematics} show kernel density estimates of the 1-D abundance, age, and kinematic distributions of all APOGEE stars, blob stars without the metal-rich contaminants, the metal-rich contaminants of the blob stars, and the metal-poor stars not allocated to the blob. The area under each distribution has been normalized to one. Interpreting these distributions should usually consider the selection function. For example, there is a bias towards seeing stars on more circular orbits around the Solar Neighbourhood, as stars with larger Galactocentric distance can only be seen at the Sun if they are on more eccentric and/or inclined orbits. However as all the distributions are subject to the same overall selection function, we make some tentative comparisons here.

The metallicity distribution (Figure \ref{fig:blob1ddist_ageabundance}) of all APOGEE stars peaks around solar as the sample is dominated by thin-disc stars. The blob stars peak at a metallicity of $\sim -1.4$, but have a wide range of metallicities. The metal-poor stars in APOGEE not in the blob have a similarly extended tail of metal-poor stars, but peak at higher metallicities. The metal-rich blob contaminants peak at the minimum metallicity of -0.8, and have a secondary peak at metallicities just below solar. 

$\aFe$ of all APOGEE stars peaks around solar, because of the dominance of the $\alpha$-poor thin-disc stars. There is a secondary peak around 0.2, which is likely to correspond to thick-disc stars. The blob stars have a peak just above 0.2, with a wide range of $\alpha$ abundances. The width of the $\aFe$ distribution of metal-poor stars in APOGEE is similar to the blob stars but it peaks at slightly higher $\aFe$ values. The metal-rich blob contaminants peaks just below 0.1, implying that they are dominated by thin-disc stars.

$\MgMn$ behaves in a similar way to $\aFe$, but the range for blob stars and metal-poor APOGEE stars is narrower than in the case for $\aFe$. This may be a result of a smaller dispersion in Mn abundances, which is typically a purer tracer of SN1a than iron. The metal-rich blob contaminants almost all have lower values compared to the rest of the blob stars, and are therefore likely to belong to the other low-$\AlFe$ component shown in Figure \ref{fig:abundances}.

$\AlFe$ is lower for the blob stars than the APOGEE stars as a whole, with a very small region of overlap. The metal-poor stars not belonging to the blob have two peaks that coincide with the other two, suggesting that it is a mixture of metal-poor disc stars, and possibly blob-like stars.

The age distribution for all APOGEE stars peaks towards younger stars due to the dominance of thin-disc stars. The ages for blob and metal-poor stars peak around 11 Gyr, with most stars between 8 to 13 Gyr. The metal-poor stars not belonging to the blob have a very similar distribution in ages, but the metal-rich blob contaminants have an age distribution more similar to the whole APOGEE sample, which is dominated by thin-disc stars. In the age distribution of blob stars, 6 stars appear to be young (i.e. < 7 Gyr). We will examine them in more detail in Section \ref{sec:discuss}.

Figure \ref{fig:blob1ddist_kinematics} shows that APOGEE stars are found primarily within about 10 kpc of the Sun, due to the selection function in distance. This is similar to the distance range examined in \cite{belokurov+18}. The distribution of blob stars and other metal-poor stars are very similar in $R$, $z$, and $s$. They are more extended in Galactocentric radius and tend to be found at higher absolute $z$. The bias towards positive values of $z$ reflects the location of the spectroscopic fields in APOGEE DR14. As blob stars are typically found at larger vertical heights, there is a bias towards seeing them with larger $J_z$. Their distances from the Sun extend out to about 20 kpc, but most are within 10 kpc.

The $z$-component of angular momentum for all APOGEE stars peaks at around 2000 km/s kpc, i.e. they are dominated by disc stars. The blob stars peak at slightly negative values of angular momentum, but cover the full extent of angular momenta of APOGEE stars. The metal-poor stars in APOGEE cover a similar range to the blob stars, but with a primary peak around zero and a secondary peak closer to 2000. This implies that it is a mixture of stars with blob-like kinematics, and disc-like kinematics. The metal-rich blob contaminants have a tail towards more positive values, further verifying the disc contamination.

The radial actions are significantly higher for the blob stars and metal-poor stars not in the blob, compared to all APOGEE stars, particularly in blob stars without the metal-rich contaminants. Therefore these stars make greater radial excursions. The blob stars without the metal-rich contaminants also appear to have a narrower distribution in $J_r$. The vertical actions are significantly higher for the blob stars compared to the other metal-poor stars and all APOGEE stars. Therefore these stars make greater vertical excursions. This again supports that the metal-rich blob contaminants are likely to be primarily disc stars that are on orbits closer to the Galactic plane.

\section{Discussion}\label{sec:discuss}
Here we discuss the ability to select purely accreted stars in the proposed abundance plane and examine the evidence for the blob stars belonging to a single progenitor.

\subsection{Ability to identify purely accreted stars}
The distribution of Milky Way stars in the $\MgFe$-$\FeH$ plane in Figure~\ref{fig:abundances_ages_actions} suggests stars in the APOGEE sample follow one of three primary chemical evolution paths. Although the selection function can play a role here \citep[e.g.][]{schonrich+11}, its role should differ between different various abundance planes. The stars in the metal-poor overdensity in the $\MgFe$-$\FeH$ plane are also present in the distinct metal-poor overdensity visible in the $\AlFe$-$\FeH$ plane. The $\MgMn$-$\AlFe$ plane was found to be a powerful way to separate metal-poor, high-$\alpha$ thick-disc stars from metal-poor, high-$\alpha$ accreted stars, and metal-rich, low-$\alpha$ thin-disc stars from the metal-poor, low-$\alpha$ accreted stars. However, there is some contamination from the higher $\MgMn$ end of this second contaminant in the selection of blob stars. About half of the contaminants are co-rotating with the disc. However, some (about half) are young, not co-rotating with the disk, metal-rich, and low in $\MgFe$. Overall, however, the contamination rate is likely to only be around $\sim7\%$, and this may be significantly reduced if $\FeH$ is used as a third dimension in the GMM.

There are a few stars in the blob for which we obtain ages that are smaller than 6 Gyr. Their kinematics do not support a disc origin. Two of these stars can be associated with the young and $\alpha$ enriched population identified by \cite{martig+15}, though they only probe stars metal-richer than $-1.0$. A possible explanation for the origin of these stars is the blue straggler scenario, i.e. the stars are not young but their higher masses are a result of mass transfer \citep[e.g.][]{jofre+16, izzard+18}. \cite{martig+15}, \cite{chiappini+15}, and the work presented here use red giants, and blue stragglers are notoriously blue. However, in an old population in which binaries coexist with isolated stars, it is inevitable that some of the evolving binaries will interact, and transfer mass at some point, creating overmassive stars. \cite{izzard+18} illustrated the creation of such stars with a population synthesis model that includes binaries. If we use isochrones of single stars to date these `overmassive' stars, they will appear young. From the spectroscopic point of view, such stars are indistinguishable from the rest of the parent population in terms of chemical composition \citep[e.g.][]{yong+16,matsuno+18}. \cite{jofre+16} show that some of these types of stars have variations in their line-of-sight velocities. However those that are the result of mergers between stars, or have periods that are larger than the time frame used to monitor the line-of-sight velocities, would not show these variations.

\subsection{Do the blob stars originate from a single progenitor or multiple progenitors?}
Several authors in the literature \citep[e.g.][]{helmi+18,belokurov+18,mackereth+19} claim that the metal-rich halo stars belong to a single progenitor that merged with the Milky Way. In reality, however, it is difficult to disentangle a single-progenitor scenario from a multiple-progenitor scenario in which there are multiple progenitors of similar masses that will suffer similarly from dynamical friction. The multiple progenitors would therefore have very similar abundances and kinematics in the Milky Way halo, but may have different SFH, which would lead to significant scatter in the $\MgFe$-$\FeH$-age relations. One argument against the multiple-progenitor scenario, is that not many massive systems could fit into the halo of the Milky Way. Another, is that simulations only predict two satellites as large as the Large Magellanic Cloud (LMC) for 6.6\% of Milky-Way sized hosts \citep{rodriguezpuebla+13}. 

Firstly we  examine the abundance and age distributions of the blob stars and then the kinematic distributions to discover what can be learned about the merger history.
\begin{table*}
 \centering
  \caption{Properties of the young blob stars. Highlighted rows show the young, high-$\alpha$ candidates. \label{tab:youngblob}}
  \begin{tabular}{llllllll}
  	\hline
  	&APOGEE ID &$V_{\mathrm{los}}$ error (km/s) &Age (Gyr) &$\FeH$ &$\aFe$ &$\AlFe$ &$L_z$ (km/s kpc)\\
  	\hline
  	1 &2M02091357+1446138 &0.020 &$2.8\pm1.2$ &$-0.95\pm0.06$ &$0.12\pm0.02$ &$-0.36\pm0.08$ &-288\\ 
  	\rowcolor{lightgray} 
  	2 &2M11515780-0147375 &0.021 &$2.6\pm0.4$ &$-1.40\pm0.05$ &$0.21\pm0.03$ &$-0.30\pm0.06$ &-185\\   	  	\rowcolor{lightgray} 
  	3 &2M13423140+2822543 &0.026 &$4.9\pm2.7$ &$-1.41\pm0.05$ &$0.23\pm0.03$ &$-0.21\pm0.06$ &748\\ 
  	4 &2M16404349+4255040 &0.012 &$5.1\pm1.8$ &$-0.83\pm0.06$ &$0.05\pm0.02$ &$-0.32\pm0.08$ &-231\\   	  	
    \hline
  \end{tabular}
\end{table*}

\subsubsection{The abundance and age distributions}
The distribution of stars in the $\MgFe$-$\FeH$-age planes supports a single progenitor origin, i.e. the oldest stars are metal poor and $\alpha$ rich, and younger stars become progressively metal richer and poorer in $\alpha$. The SFH cannot be directly inferred from the age distribution due to a selection bias in the age-metallicity distribution, as a result of the colour-magnitude selection criteria for APOGEE stars. However the range implies that star formation occurred over approximately 5 Gyr, and that a single progenitor would have merged with the Milky Way about 8 Gyr ago. The SFH must be rather different from the LMC, which is similar in mass. The LMC is likely to have had a recent burst of star formation \citep{sanders+18,nidever+19}, which leads to an increase in $\aFe$ between a metallicity of -1.5 and -1.0. 

\begin{figure}
    \centering
    \includegraphics[scale=0.6]{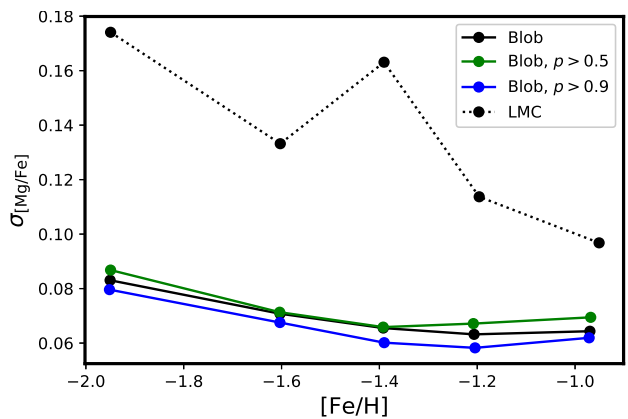}
    \caption{The scatter (as measured by the standard deviation) in $\MgFe$ as a function of $\FeH$, for the blob stars (solid) for different probability cut-offs for belonging to the blob, and stars belonging to the LMC (dotted).}
    \label{fig:abundancescatter}
\end{figure}

Multiple progenitors could create a similar distribution in the $\MgFe$-$\FeH$-age plane, but they would have to have similar SFH. Multiple progenitors are also likely to lead to an increase in the scatter in these relations. In Figure \ref{fig:abundancescatter}, we compare the scatter in the $\MgFe$-$\FeH$ for our blob stars to that for stars in the LMC. The LMC is likely to be of a very similar total mass compared to the single progenitor associated with the blob (see Section \ref{sssec:discuss_blob_kinematics}). We estimate the scatter for the LMC stars with a signal-to-noise ratio above 40, by estimating the chemical abundances directly from Figure 9 in \cite{nidever+19}. We estimate the scatter for the blob stars using our sample. As imposing a cut-off probability for assigning stars to a Gaussian component might reduce the scatter inferred, other potential samples of blob stars assuming a probability of belonging to the Gaussian component of 0.5 and 0.9 were also examined.

The blob stars all have a scatter between around 0.06 and 0.08, which decreases slightly with metallicity. The LMC stars have a scatter that approximately decreases from 0.18 to 0.10 over the same range in metallicity. The decrease with metallicity reflects both how abundances are measured better in metal-richer stars, and the effect of inhomogeneous enrichment on metal-poorer stars. As a comparison, the mean uncertainty in $\MgFe$ for our sample of APOGEE stars is 0.02. Some of this discrepancy must arise from the different ranges in both samples, e.g. in signal-to-noise, abundance uncertainty, and effective temperature. This does entertain the possibility of the blob stars belonging to a single progenitor, as otherwise one would have expected a comparable scatter to the LMC. 

\subsubsection{Kinematic distributions}
\label{sssec:discuss_blob_kinematics}
The blob stars are mainly found within 10 kpc, though many are as far as 20 kpc. Like other metal-poor stars, they are found at higher values of radial action, which may be a result of the selection function, i.e. metal-poor stars are generally found further away. Therefore they are only observed close to the Solar Neighbourhood if they are on more radially anisotropic orbits. This can only be properly understood in the context of models. Their distribution in $J_r$ is noticeably narrower however. The width in $J_r$ can be used to make an estimate of the mass of a single progenitor \citep{eyre+11}. The radial action is given by
\begin{equation}
\label{eq:Jr}
J_r = \frac{1}{\pi}\oint v_r \mathrm{d} r,
\end{equation}
where the integration path is one complete integration. If we consider a particle whose radial velocity differs from the system's average by $\sigma$, i.e. $\delta v_r \sim \sigma_r$, then we can take a finite difference over Equation \eqref{eq:Jr}. This gives us an expression for the difference between the radial action of a particle from the system's average
\begin{equation}
\label{eq:sigrest}
\delta J_r \sim \frac{1}{\pi} \sigma_r \Delta r,
\end{equation}
where $\Delta r = R_p-R_a$ is the amplitude of the radial oscillation, $R_p$ is the pericentre of the orbit, and $R_a$ is the apocentre of the orbit. Then from the virial theorem, KE$_{\mathrm{avg}}$ = $-\frac{1}{2}$GPE$_{\mathrm{avg}}$, which gives:
\begin{equation}
\label{eq:vt}
\frac{1}{2}m(\sigma_r)^2 = \frac{1}{2}\frac{GM_bm}{\frac{1}{2}R_b},
\end{equation}
where $M_b$ is the mass of the blob, and $R_b$ is the radius of the single system from which the blob originates. Approximating the action width, $\delta J_r$, by the dispersion in $J_r$, and the radial oscillation, $\Delta r$, by the $95\%$ range in $R$, and assuming the accreted system to be a large dwarf spheroidal with a radius of $\sim 5$ kpc, Equation \eqref{eq:vt} gives a total progenitor mass of $\sim3.4\times10^{11}M_{\odot}$. This is  similar to that recently estimated for the LMC \citep{erkal+18}. Using an abundance-matching based mapping from total mass to stellar mass \citep[e.g.][]{read+17}, this implies a stellar mass $\sim 10^{9.5}M_{\odot}$. Chemical evolution models applied to the distribution of accreted stars in the $\MgFe$-$\FeH$ plane find a stellar mass of $\sim 6 \times 10^8M_{\odot}$ \citep{fernandez+18,vincenzo+19}. \cite{mackereth+19} compare the distribution of the accreted stars in the $\MgFe$-$\FeH$ plane with those of accreted galaxies from the EAGLE suite of cosmological simulations, and constrain the stellar mass of a single progenitor to be between $10^{8.5}M_{\odot}$ and $10^9M_{\odot}$. 

\section{Conclusions}
We use a GMM to select a `blob' of accreted stars in the $\MgMn$-$\AlFe$ abundance plane, using abundances published in APOGEE DR14. The blob stars are found to span a range of metallicities from -0.5 to -2.5 and $\MgFe$ from -0.1 to 0.5. They constitute $\sim$30\% of the metal-poor ($\FeH < -0.8$) halo at metallicities of $\sim-1.4$. The ages from \cite{sanders+18} imply a significant fraction of young ages, potentially as a result of the effects of thermohaline mixing and stochastic chemical evolution at lower metallicities. Our new ages are found to mainly range from 8 to 13 Gyr, with the oldest stars the metal-poorest, and with the highest $\MgFe$ abundance. The blob stars are mainly found within 10 kpc, though many are as far as 20 kpc. Like other metal-poor stars, they are found at higher values of radial action, but their distribution in radial action is noticeably narrower. Their metallicities and distances from the Sun imply that the sample is similar to that examined from the SDSS-Gaia sample in \cite{belokurov+18}.

The blob stars exhibit the expected age, metallicity, and $\alpha$ sequence for stars belonging to a single system, however the variation in scatter in $\MgFe$ against $\FeH$ appears to be higher than stars observed with APOGEE belonging to the LMC, which is predicted to have a very similar mass to such a single system. The blob stars tend to make larger radial excursions compared to the average APOGEE star. \cite{belokurov+18} argue that the high level of radial anisotropy for `blob' stars points towards a massive progenitor whose stellar orbits have been heavily radialized through a combination of dynamical friction and disc formation. We note, that all metal-poor stars tend to be more radially anisotropic as a result of a selection effect. The $J_R$ distribution for blob stars appear to be more narrow however. Dynamical arguments analyzing this distribution suggest a single system with a total mass of $\sim10^{11}M_{\odot}$, similar to that found by other authors in the literature. Finally, we find 4 young (i.e. < 6 Gyr) stars in the blob. Two of these appear to extend the young $\alpha$-enriched stars of \cite{martig+15} to lower metallicities. 

APOGEE includes elements from a range of nucleosynthesis processes that drive chemical evolution in the Galaxy, but is absent in heavy, neutron-rich elements ($Z \ge 30$), which are mainly produced by neutron capture via both the s- and r-processes. There are abundances for one heavy element (Rb; $Z = 37$) but these results are based upon very weak lines of Rb I. The s-process element cerium has a high density of absorption lines expected in the spectrum of red giant stars and consists primarily of two stable isotopes. The s-process accounts for $\sim$90\% of it, with the r-process contributing the other $\sim$10\%. This element may become available in future data releases \citep{cunha+17}, and spectra of these stars in the optical part of the spectrum will also help reveal more s-process elements. Using these in the GMM procedure will better characterize the blob stars. Over the next few months we will be observing several of these stars in the optical part of the spectrum, giving access to several s-process and r-process elements.

In the future, we plan to formally disentangle a single from a multiple-progenitor scenario using chemical evolution models of the type ran recently by \cite{vincenzo+19}, taking the age-metallicity selection function into account. 

\section*{Acknowledgements}
PD would like to acknowledge support from the STFC (ST/N000919/1). PJ acknowledges support of FONDECYT Iniciaci\'on Grant Number 11170174.  This work has made use of data from the European Space Agency (ESA) mission {\it Gaia} (\url{https://www.cosmos.esa.int/gaia}), processed by the {\it Gaia} Data Processing and Analysis Consortium (DPAC,
\url{https://www.cosmos.esa.int/web/gaia/dpac/consortium}). Funding for the DPAC has been provided by national institutions, in particular the institutions participating in the {\it Gaia} Multilateral Agreement.

Funding for the Sloan Digital Sky Survey IV has been provided by the Alfred P. Sloan Foundation, the U.S. Department of Energy Office of Science, and the Participating Institutions. SDSS-IV acknowledges support and resources from the Center for High-Performance Computing at the University of Utah. The SDSS web site is www.sdss.org.




\bibliographystyle{mnras}
\bibliography{bibliography} 

\begin{thebibliography}{}
\makeatletter
\relax
\def\mn@urlcharsother{\let\do\@makeother \do\$\do\&\do\#\do\^\do\_\do\%\do\~}
\def\mn@doi{\begingroup\mn@urlcharsother \@ifnextchar [ {\mn@doi@}
  {\mn@doi@[]}}
\def\mn@doi@[#1]#2{\def\@tempa{#1}\ifx\@tempa\@empty \href
  {http://dx.doi.org/#2} {doi:#2}\else \href {http://dx.doi.org/#2} {#1}\fi
  \endgroup}
\def\mn@eprint#1#2{\mn@eprint@#1:#2::\@nil}
\def\mn@eprint@arXiv#1{\href {http://arxiv.org/abs/#1} {{\tt arXiv:#1}}}
\def\mn@eprint@dblp#1{\href {http://dblp.uni-trier.de/rec/bibtex/#1.xml}
  {dblp:#1}}
\def\mn@eprint@#1:#2:#3:#4\@nil{\def\@tempa {#1}\def\@tempb {#2}\def\@tempc
  {#3}\ifx \@tempc \@empty \let \@tempc \@tempb \let \@tempb \@tempa \fi \ifx
  \@tempb \@empty \def\@tempb {arXiv}\fi \@ifundefined
  {mn@eprint@\@tempb}{\@tempb:\@tempc}{\expandafter \expandafter \csname
  mn@eprint@\@tempb\endcsname \expandafter{\@tempc}}}

\bibitem[\protect\citeauthoryear{{Abadi}, {Navarro}  \& {Steinmetz}}{{Abadi}
  et~al.}{2006}]{abadi+06}
{Abadi} M.~G.,  {Navarro} J.~F.,   {Steinmetz} M.,  2006, \mn@doi [\mnras]
  {10.1111/j.1365-2966.2005.09789.x}, 365, 747

\bibitem[\protect\citeauthoryear{{Abolfathi} et~al.,}{{Abolfathi}
  et~al.}{2018}]{abol+18}
{Abolfathi} B.,  et~al., 2018, \mn@doi [\apjs] {10.3847/1538-4365/aa9e8a},
  \href {http://adsabs.harvard.edu/abs/2018ApJS..235...42A} {235, 42}

\bibitem[\protect\citeauthoryear{{Allende Prieto} et~al.,}{{Allende Prieto}
  et~al.}{2014}]{allende+14}
{Allende Prieto} C.,  et~al., 2014, \mn@doi [\aap]
  {10.1051/0004-6361/201424053}, \href
  {http://adsabs.harvard.edu/abs/2014A%26A...568A...7A} {568, A7}

\bibitem[\protect\citeauthoryear{{Bell} et~al.,}{{Bell} et~al.}{2008}]{bell+08}
{Bell} E.~F.,  et~al., 2008, \mn@doi [\apj] {10.1086/588032}, 680, 295

\bibitem[\protect\citeauthoryear{{Bell}, {Xue}, {Rix}, {Ruhland}  \&
  {Hogg}}{{Bell} et~al.}{2010}]{bell+10}
{Bell} E.~F.,  {Xue} X.~X.,  {Rix} H.-W.,  {Ruhland} C.,   {Hogg} D.~W.,  2010,
  \mn@doi [\aj] {10.1088/0004-6256/140/6/1850}, \href
  {http://adsabs.harvard.edu/abs/2010AJ....140.1850B} {140, 1850}

\bibitem[\protect\citeauthoryear{{Belokurov} et~al.,}{{Belokurov}
  et~al.}{2007}]{belokurov+07}
{Belokurov} V.,  et~al., 2007, \mn@doi [\apj] {10.1086/509718}, 654, 897

\bibitem[\protect\citeauthoryear{{Belokurov}, {Erkal}, {Evans}, {Koposov}  \&
  {Deason}}{{Belokurov} et~al.}{2018}]{belokurov+18}
{Belokurov} V.,  {Erkal} D.,  {Evans} N.~W.,  {Koposov} S.~E.,   {Deason}
  A.~J.,  2018, \mn@doi [\mnras] {10.1093/mnras/sty982}, \href
  {https://ui.adsabs.harvard.edu/\#abs/2018MNRAS.478..611B} {478, 611}

\bibitem[\protect\citeauthoryear{{Bensby}, {Feltzing}  \& {Oey}}{{Bensby}
  et~al.}{2014}]{bensby+14}
{Bensby} T.,  {Feltzing} S.,   {Oey} M.~S.,  2014, \mn@doi [\aap]
  {10.1051/0004-6361/201322631}, \href
  {https://ui.adsabs.harvard.edu/\#abs/2014A&A...562A..71B} {562, A71}

\bibitem[\protect\citeauthoryear{{Bland-Hawthorn} \&
  {Gerhard}}{{Bland-Hawthorn} \& {Gerhard}}{2016}]{bland+16}
{Bland-Hawthorn} J.,  {Gerhard} O.,  2016, \mn@doi [\araa]
  {10.1146/annurev-astro-081915-023441}, \href
  {http://adsabs.harvard.edu/abs/2016ARA%26A..54..529B} {54, 529}

\bibitem[\protect\citeauthoryear{{Bond} et~al.,}{{Bond} et~al.}{2010}]{bond+10}
{Bond} N.~A.,  et~al., 2010, \mn@doi [\apj] {10.1088/0004-637X/716/1/1}, 716, 1

\bibitem[\protect\citeauthoryear{{Bovy}}{{Bovy}}{2017}]{bovy17}
{Bovy} J.,  2017, \mn@doi [\mnras] {10.1093/mnras/stx1277}, \href
  {http://adsabs.harvard.edu/abs/2017MNRAS.470.1360B} {470, 1360}

\bibitem[\protect\citeauthoryear{{Bovy}, {Rix}, {Green}, {Schlafly}  \&
  {Finkbeiner}}{{Bovy} et~al.}{2016}]{bovy+16b}
{Bovy} J.,  {Rix} H.-W.,  {Green} G.~M.,  {Schlafly} E.~F.,   {Finkbeiner}
  D.~P.,  2016, \mn@doi [\apj] {10.3847/0004-637X/818/2/130}, \href
  {http://adsabs.harvard.edu/abs/2016ApJ...818..130B} {818, 130}

\bibitem[\protect\citeauthoryear{{Bressan}, {Marigo}, {Girardi}, {Salasnich},
  {Dal Cero}, {Rubele}  \& {Nanni}}{{Bressan} et~al.}{2012}]{bressan+12}
{Bressan} A.,  {Marigo} P.,  {Girardi} L.,  {Salasnich} B.,  {Dal Cero} C.,
  {Rubele} S.,   {Nanni} A.,  2012, \mn@doi [\mnras]
  {10.1111/j.1365-2966.2012.21948.x}, \href
  {http://adsabs.harvard.edu/abs/2012MNRAS.427..127B} {427, 127}

\bibitem[\protect\citeauthoryear{{Burnett} \& {Binney}}{{Burnett} \&
  {Binney}}{2010}]{burnett+10}
{Burnett} B.,  {Binney} J.,  2010, \mn@doi [\mnras]
  {10.1111/j.1365-2966.2010.16896.x}, \href
  {http://adsabs.harvard.edu/abs/2010MNRAS.407..339B} {407, 339}

\bibitem[\protect\citeauthoryear{{Carollo} et~al.,}{{Carollo}
  et~al.}{2007}]{carollo+07}
{Carollo} D.,  et~al., 2007, \mn@doi [\nat] {10.1038/nature06460}, 450, 1020

\bibitem[\protect\citeauthoryear{{Chen}, {Zhao}, {Carrell}, {Zhao}, {Tan},
  {Nissen}  \& {Wei}}{{Chen} et~al.}{2014}]{chen+14}
{Chen} Y.~Q.,  {Zhao} G.,  {Carrell} K.,  {Zhao} J.~K.,  {Tan} K.~F.,  {Nissen}
  P.~E.,   {Wei} P.,  2014, \mn@doi [\apj] {10.1088/0004-637X/795/1/52}, \href
  {http://adsabs.harvard.edu/abs/2014ApJ...795...52C} {795, 52}

\bibitem[\protect\citeauthoryear{{Chiappini} et~al.,}{{Chiappini}
  et~al.}{2015}]{chiappini+15}
{Chiappini} C.,  et~al., 2015, \mn@doi [\aap] {10.1051/0004-6361/201525865},
  \href {http://adsabs.harvard.edu/abs/2015A%26A...576L..12C} {576, L12}

\bibitem[\protect\citeauthoryear{{Chiba} \& {Yoshii}}{{Chiba} \&
  {Yoshii}}{1998}]{chiba+98}
{Chiba} M.,  {Yoshii} Y.,  1998, \mn@doi [\aj] {10.1086/300177}, \href
  {https://ui.adsabs.harvard.edu/\#abs/1998AJ....115..168C} {115, 168}

\bibitem[\protect\citeauthoryear{{Cooper}, {Cole}, {Frenk}  \&
  {Helmi}}{{Cooper} et~al.}{2011}]{cooper+11}
{Cooper} A.~P.,  {Cole} S.,  {Frenk} C.~S.,   {Helmi} A.,  2011, \mn@doi
  [\mnras] {10.1111/j.1365-2966.2011.19401.x}, 417, 2206

\bibitem[\protect\citeauthoryear{{Cunha} et~al.,}{{Cunha}
  et~al.}{2017}]{cunha+17}
{Cunha} K.,  et~al., 2017, \mn@doi [\apj] {10.3847/1538-4357/aa7beb}, \href
  {https://ui.adsabs.harvard.edu/\#abs/2017ApJ...844..145C} {844, 145}

\bibitem[\protect\citeauthoryear{{Cunningham} et~al.,}{{Cunningham}
  et~al.}{2016}]{cunningham+16}
{Cunningham} E.~C.,  et~al., 2016, \mn@doi [\apj] {10.3847/0004-637X/820/1/18},
  \href {http://adsabs.harvard.edu/abs/2016ApJ...820...18C} {820, 18}

\bibitem[\protect\citeauthoryear{{Das} \& {Binney}}{{Das} \&
  {Binney}}{2016}]{das+16a}
{Das} P.,  {Binney} J.,  2016, \mn@doi [\mnras] {10.1093/mnras/stw744}, \href
  {http://adsabs.harvard.edu/abs/2016MNRAS.460.1725D} {460, 1725}

\bibitem[\protect\citeauthoryear{{Das} \& {Sanders}}{{Das} \&
  {Sanders}}{2019}]{das+19}
{Das} P.,  {Sanders} J.~L.,  2019, \mn@doi [\mnras] {10.1093/mnras/sty2776},
  \href {https://ui.adsabs.harvard.edu/\#abs/2019MNRAS.484..294D} {484, 294}

\bibitem[\protect\citeauthoryear{{Das}, {Williams}  \& {Binney}}{{Das}
  et~al.}{2016}]{das+16b}
{Das} P.,  {Williams} A.,   {Binney} J.,  2016, \mn@doi [\mnras]
  {10.1093/mnras/stw2167}, \href
  {http://adsabs.harvard.edu/abs/2016MNRAS.463.3169D} {463, 3169}

\bibitem[\protect\citeauthoryear{{Deason}, {Belokurov}  \& {Evans}}{{Deason}
  et~al.}{2011}]{deason+11a}
{Deason} A.~J.,  {Belokurov} V.,   {Evans} N.~W.,  2011, \mn@doi [\mnras]
  {10.1111/j.1365-2966.2011.19237.x}, 416, 2903

\bibitem[\protect\citeauthoryear{{Deason}, {Belokurov}, {Evans}  \&
  {An}}{{Deason} et~al.}{2012}]{deason+12a}
{Deason} A.~J.,  {Belokurov} V.,  {Evans} N.~W.,   {An} J.,  2012, \mn@doi
  [\mnras] {10.1111/j.1745-3933.2012.01283.x}, 424, L44

\bibitem[\protect\citeauthoryear{{Deason}, {Belokurov}, {Evans}  \&
  {Johnston}}{{Deason} et~al.}{2013a}]{deason+13a}
{Deason} A.~J.,  {Belokurov} V.,  {Evans} N.~W.,   {Johnston} K.~V.,  2013a,
  \mn@doi [\apj] {10.1088/0004-637X/763/2/113}, \href
  {https://ui.adsabs.harvard.edu/#abs/2013ApJ...763..113D} {763, 113}

\bibitem[\protect\citeauthoryear{{Deason}, {Van der Marel}, {Guhathakurta},
  {Sohn}  \& {Brown}}{{Deason} et~al.}{2013b}]{deason+13b}
{Deason} A.~J.,  {Van der Marel} R.~P.,  {Guhathakurta} P.,  {Sohn} S.~T.,
  {Brown} T.~M.,  2013b, \mn@doi [\apj] {10.1088/0004-637X/766/1/24}, \href
  {http://adsabs.harvard.edu/abs/2013ApJ...766...24D} {766, 24}

\bibitem[\protect\citeauthoryear{{Deason}, {Belokurov}, {Koposov}  \&
  {Rockosi}}{{Deason} et~al.}{2014}]{deason+14}
{Deason} A.~J.,  {Belokurov} V.,  {Koposov} S.~E.,   {Rockosi} C.~M.,  2014,
  \mn@doi [\apj] {10.1088/0004-637X/787/1/30}, \href
  {http://adsabs.harvard.edu/abs/2014ApJ...787...30D} {787, 30}

\bibitem[\protect\citeauthoryear{{Erkal} et~al.,}{{Erkal}
  et~al.}{2018}]{erkal+18}
{Erkal} D.,  et~al., 2018, arXiv e-prints, \href
  {https://ui.adsabs.harvard.edu/\#abs/2018arXiv181208192E} {p.
  arXiv:1812.08192}

\bibitem[\protect\citeauthoryear{{Eyre} \& {Binney}}{{Eyre} \&
  {Binney}}{2011}]{eyre+11}
{Eyre} A.,  {Binney} J.,  2011, \mn@doi [\mnras]
  {10.1111/j.1365-2966.2011.18270.x}, \href
  {https://ui.adsabs.harvard.edu/\#abs/2011MNRAS.413.1852E} {413, 1852}

\bibitem[\protect\citeauthoryear{{Fern{\'a}ndez-Alvar}
  et~al.,}{{Fern{\'a}ndez-Alvar} et~al.}{2018}]{fernandez+18}
{Fern{\'a}ndez-Alvar} E.,  et~al., 2018, \mn@doi [\apj]
  {10.3847/1538-4357/aa9ced}, \href
  {https://ui.adsabs.harvard.edu/\#abs/2018ApJ...852...50F} {852, 50}

\bibitem[\protect\citeauthoryear{{Font}, {McCarthy}, {Crain}, {Theuns},
  {Schaye}, {Wiersma}  \& {Dalla Vecchia}}{{Font} et~al.}{2011}]{font+11}
{Font} A.~S.,  {McCarthy} I.~G.,  {Crain} R.~A.,  {Theuns} T.,  {Schaye} J.,
  {Wiersma} R.~P.~C.,   {Dalla Vecchia} C.,  2011, \mn@doi [\mnras]
  {10.1111/j.1365-2966.2011.19227.x}, \href
  {https://ui.adsabs.harvard.edu/\#abs/2011MNRAS.416.2802F} {416, 2802}

\bibitem[\protect\citeauthoryear{{Gaia Collaboration}, {Brown}, {Vallenari},
  {Prusti}, {de Bruijne}, {Babusiaux}  \& {Bailer-Jones}}{{Gaia Collaboration}
  et~al.}{2018}]{gaia+18}
{Gaia Collaboration} {Brown} A.~G.~A.,  {Vallenari} A.,  {Prusti} T.,  {de
  Bruijne} J.~H.~J.,  {Babusiaux} C.,   {Bailer-Jones} C.~A.~L.,  2018,
  preprint, \href {http://adsabs.harvard.edu/abs/2018arXiv180409365G} {}
  (\mn@eprint {arXiv} {1804.09365})

\bibitem[\protect\citeauthoryear{{Garc{\'{\i}}a P{\'e}rez}
  et~al.,}{{Garc{\'{\i}}a P{\'e}rez} et~al.}{2016}]{garcia+16}
{Garc{\'{\i}}a P{\'e}rez} A.~E.,  et~al., 2016, \mn@doi [\aj]
  {10.3847/0004-6256/151/6/144}, \href
  {http://adsabs.harvard.edu/abs/2016AJ....151..144G} {151, 144}

\bibitem[\protect\citeauthoryear{{Haas} et~al.,}{{Haas} et~al.}{2010}]{haas+10}
{Haas} M.~R.,  et~al., 2010, \mn@doi [\apjl] {10.1088/2041-8205/713/2/L115},
  \href {http://adsabs.harvard.edu/abs/2010ApJ...713L.115H} {713, L115}

\bibitem[\protect\citeauthoryear{{Hattori}, {Valluri}, {Loebman}  \&
  {Bell}}{{Hattori} et~al.}{2017}]{hattori+17}
{Hattori} K.,  {Valluri} M.,  {Loebman} S.~R.,   {Bell} E.~F.,  2017, \mn@doi
  [\apj] {10.3847/1538-4357/aa71aa}, \href
  {https://ui.adsabs.harvard.edu/\#abs/2017ApJ...841...91H} {841, 91}

\bibitem[\protect\citeauthoryear{{Hawkins}, {Jofr{\'e}}, {Gilmore}  \&
  {Masseron}}{{Hawkins} et~al.}{2014}]{hawkins+14}
{Hawkins} K.,  {Jofr{\'e}} P.,  {Gilmore} G.,   {Masseron} T.,  2014, \mn@doi
  [\mnras] {10.1093/mnras/stu1910}, \href
  {http://adsabs.harvard.edu/abs/2014MNRAS.445.2575H} {445, 2575}

\bibitem[\protect\citeauthoryear{{Hawkins}, {Jofr{\'e}}, {Masseron}  \&
  {Gilmore}}{{Hawkins} et~al.}{2015}]{hawkins+15}
{Hawkins} K.,  {Jofr{\'e}} P.,  {Masseron} T.,   {Gilmore} G.,  2015, \mn@doi
  [\mnras] {10.1093/mnras/stv1586}, \href
  {https://ui.adsabs.harvard.edu/#abs/2015MNRAS.453..758H} {453, 758}

\bibitem[\protect\citeauthoryear{{Hayden} et~al.,}{{Hayden}
  et~al.}{2014}]{hayden+14}
{Hayden} M.~R.,  et~al., 2014, \mn@doi [\aj] {10.1088/0004-6256/147/5/116},
  \href {http://adsabs.harvard.edu/abs/2014AJ....147..116H} {147, 116}

\bibitem[\protect\citeauthoryear{{Hayes} et~al.,}{{Hayes}
  et~al.}{2018}]{hayes+18}
{Hayes} C.~R.,  et~al., 2018, \mn@doi [\apj] {10.3847/1538-4357/aa9cec}, \href
  {https://ui.adsabs.harvard.edu/\#abs/2018ApJ...852...49H} {852, 49}

\bibitem[\protect\citeauthoryear{Haywood, Matteo, Lehnert, Snaith, Khoperskov
  \& Gómez}{Haywood et~al.}{2018}]{haywood+18}
Haywood M.,  Matteo P.~D.,  Lehnert M.~D.,  Snaith O.,  Khoperskov S.,   Gómez
  A.,  2018, The Astrophysical Journal, 863, 113

\bibitem[\protect\citeauthoryear{{Helmi}, {Babusiaux}, {Koppelman}, {Massari},
  {Veljanoski}  \& {Brown}}{{Helmi} et~al.}{2018}]{helmi+18}
{Helmi} A.,  {Babusiaux} C.,  {Koppelman} H.~H.,  {Massari} D.,  {Veljanoski}
  J.,   {Brown} A. G.~A.,  2018, \mn@doi [\nat] {10.1038/s41586-018-0625-x},
  \href {https://ui.adsabs.harvard.edu/#abs/2018Natur.563...85H} {563, 85}

\bibitem[\protect\citeauthoryear{{Holtzman} et~al.,}{{Holtzman}
  et~al.}{2018}]{holtzman+18}
{Holtzman} J.~A.,  et~al., 2018, \mn@doi [\aj] {10.3847/1538-3881/aad4f9},
  \href {https://ui.adsabs.harvard.edu/\#abs/2018AJ....156..125H} {156, 125}

\bibitem[\protect\citeauthoryear{{Ibata}, {Gilmore}  \& {Irwin}}{{Ibata}
  et~al.}{1995}]{ibata+95}
{Ibata} R.~A.,  {Gilmore} G.,   {Irwin} M.~J.,  1995, \mnras, 277, 781

\bibitem[\protect\citeauthoryear{{Ivezi{\'c}} et~al.,}{{Ivezi{\'c}}
  et~al.}{2008}]{ivezic+08}
{Ivezi{\'c}} {\v{Z}}.,  et~al., 2008, \mn@doi [\apj] {10.1086/589678}, \href
  {https://ui.adsabs.harvard.edu/\#abs/2008ApJ...684..287I} {684, 287}

\bibitem[\protect\citeauthoryear{{Izzard}, {Preece}, {Jofre}, {Halabi},
  {Masseron}  \& {Tout}}{{Izzard} et~al.}{2018}]{izzard+18}
{Izzard} R.~G.,  {Preece} H.,  {Jofre} P.,  {Halabi} G.~M.,  {Masseron} T.,
  {Tout} C.~A.,  2018, \mn@doi [\mnras] {10.1093/mnras/stx2355}, \href
  {http://adsabs.harvard.edu/abs/2018MNRAS.473.2984I} {473, 2984}

\bibitem[\protect\citeauthoryear{{Jofr{\'e}} \& {Weiss}}{{Jofr{\'e}} \&
  {Weiss}}{2011}]{jofre+11}
{Jofr{\'e}} P.,  {Weiss} A.,  2011, \mn@doi [\aap]
  {10.1051/0004-6361/201117131}, 533, A59

\bibitem[\protect\citeauthoryear{{Jofr{\'e}} et~al.,}{{Jofr{\'e}}
  et~al.}{2016}]{jofre+16}
{Jofr{\'e}} P.,  et~al., 2016, \mn@doi [\aap] {10.1051/0004-6361/201629356},
  \href {http://adsabs.harvard.edu/abs/2016A%26A...595A..60J} {595, A60}

\bibitem[\protect\citeauthoryear{{Jofr{\'e}}, {Heiter}  \&
  {Soubiran}}{{Jofr{\'e}} et~al.}{2018}]{jofre+18}
{Jofr{\'e}} P.,  {Heiter} U.,   {Soubiran} C.,  2018, arXiv e-prints, \href
  {http://adsabs.harvard.edu/abs/2018arXiv181108041J} {}

\bibitem[\protect\citeauthoryear{{Kafle}, {Sharma}, {Lewis}  \&
  {Bland-Hawthorn}}{{Kafle} et~al.}{2012}]{kafle+12}
{Kafle} P.~R.,  {Sharma} S.,  {Lewis} G.~F.,   {Bland-Hawthorn} J.,  2012,
  \mn@doi [\apj] {10.1088/0004-637X/761/2/98}, \href
  {http://adsabs.harvard.edu/abs/2012ApJ...761...98K} {761, 98}

\bibitem[\protect\citeauthoryear{{Kafle}, {Sharma}, {Lewis}  \&
  {Bland-Hawthorn}}{{Kafle} et~al.}{2013}]{kafle+13}
{Kafle} P.~R.,  {Sharma} S.,  {Lewis} G.~F.,   {Bland-Hawthorn} J.,  2013,
  \mn@doi [\mnras] {10.1093/mnras/stt101}, 430, 2973

\bibitem[\protect\citeauthoryear{{Kalirai}}{{Kalirai}}{2012}]{kali+12}
{Kalirai} J.~S.,  2012, \mn@doi [\nat] {10.1038/nature11062}, \href
  {http://adsabs.harvard.edu/abs/2012Natur.486...90K} {486, 90}

\bibitem[\protect\citeauthoryear{{Kobayashi}, {Umeda}, {Nomoto}, {Tominaga}  \&
  {Ohkubo}}{{Kobayashi} et~al.}{2006}]{kobayashi+06}
{Kobayashi} C.,  {Umeda} H.,  {Nomoto} K.,  {Tominaga} N.,   {Ohkubo} T.,
  2006, \mn@doi [\apj] {10.1086/508914}, \href
  {https://ui.adsabs.harvard.edu/\#abs/2006ApJ...653.1145K} {653, 1145}

\bibitem[\protect\citeauthoryear{{Kroupa}, {Tout}  \& {Gilmore}}{{Kroupa}
  et~al.}{1993}]{kroupa+93}
{Kroupa} P.,  {Tout} C.~A.,   {Gilmore} G.,  1993, \mnras, 262, 545

\bibitem[\protect\citeauthoryear{{Lagarde} et~al.,}{{Lagarde}
  et~al.}{2018}]{lagarde+18}
{Lagarde} N.,  et~al., 2018, arXiv e-prints, \href
  {https://ui.adsabs.harvard.edu/\#abs/2018arXiv180601868L} {p.
  arXiv:1806.01868}

\bibitem[\protect\citeauthoryear{{Lindegren} et~al.,}{{Lindegren}
  et~al.}{2016}]{lindegren+16}
{Lindegren} L.,  et~al., 2016, \mn@doi [\aap] {10.1051/0004-6361/201628714},
  \href {https://ui.adsabs.harvard.edu/#abs/2016A&A...595A...4L} {595, A4}

\bibitem[\protect\citeauthoryear{{Lindegren} et~al.,}{{Lindegren}
  et~al.}{2018}]{lindegren+18}
{Lindegren} L.,  et~al., 2018, \mn@doi [\aap] {10.1051/0004-6361/201832727},
  \href {https://ui.adsabs.harvard.edu/\#abs/2018A&A...616A...2L} {616, A2}

\bibitem[\protect\citeauthoryear{{Mackereth} et~al.,}{{Mackereth}
  et~al.}{2019}]{mackereth+19}
{Mackereth} J.~T.,  et~al., 2019, \mn@doi [\mnras] {10.1093/mnras/sty2955},
  \href {https://ui.adsabs.harvard.edu/\#abs/2019MNRAS.482.3426M} {482, 3426}

\bibitem[\protect\citeauthoryear{{Majewski}, {Skrutskie}, {Weinberg}  \&
  {Ostheimer}}{{Majewski} et~al.}{2003}]{majewski+03}
{Majewski} S.~R.,  {Skrutskie} M.~F.,  {Weinberg} M.~D.,   {Ostheimer} J.~C.,
  2003, \mn@doi [\apj] {10.1086/379504}, 599, 1082

\bibitem[\protect\citeauthoryear{{Martig} et~al.,}{{Martig}
  et~al.}{2015}]{martig+15}
{Martig} M.,  et~al., 2015, \mn@doi [\mnras] {10.1093/mnras/stv1071}, \href
  {https://ui.adsabs.harvard.edu/\#abs/2015MNRAS.451.2230M} {451, 2230}

\bibitem[\protect\citeauthoryear{{Martig} et~al.,}{{Martig}
  et~al.}{2016}]{martig+16}
{Martig} M.,  et~al., 2016, \mn@doi [\mnras] {10.1093/mnras/stv2830}, \href
  {http://adsabs.harvard.edu/abs/2016MNRAS.456.3655M} {456, 3655}

\bibitem[\protect\citeauthoryear{{Masseron} \& {Gilmore}}{{Masseron} \&
  {Gilmore}}{2015}]{masseron+15}
{Masseron} T.,  {Gilmore} G.,  2015, \mn@doi [\mnras] {10.1093/mnras/stv1731},
  \href {http://adsabs.harvard.edu/abs/2015MNRAS.453.1855M} {453, 1855}

\bibitem[\protect\citeauthoryear{{Matsuno}, {Yong}, {Aoki}  \&
  {Ishigaki}}{{Matsuno} et~al.}{2018}]{matsuno+18}
{Matsuno} T.,  {Yong} D.,  {Aoki} W.,   {Ishigaki} M.~N.,  2018, \mn@doi [\apj]
  {10.3847/1538-4357/aac019}, \href
  {http://adsabs.harvard.edu/abs/2018ApJ...860...49M} {860, 49}

\bibitem[\protect\citeauthoryear{{Myeong}, {Evans}, {Belokurov}, {Sanders}  \&
  {Koposov}}{{Myeong} et~al.}{2018}]{myeong+18}
{Myeong} G.~C.,  {Evans} N.~W.,  {Belokurov} V.,  {Sanders} J.~L.,   {Koposov}
  S.~E.,  2018, \mn@doi [\apj] {10.3847/2041-8213/aad7f7}, \href
  {https://ui.adsabs.harvard.edu/\#abs/2018ApJ...863L..28M} {863, L28}

\bibitem[\protect\citeauthoryear{{Ness}, {Hogg}, {Rix}, {Martig},
  {Pinsonneault}  \& {Ho}}{{Ness} et~al.}{2016}]{ness+16}
{Ness} M.,  {Hogg} D.~W.,  {Rix} H.-W.,  {Martig} M.,  {Pinsonneault} M.~H.,
  {Ho} A.~Y.~Q.,  2016, \mn@doi [\apj] {10.3847/0004-637X/823/2/114}, \href
  {http://adsabs.harvard.edu/abs/2016ApJ...823..114N} {823, 114}

\bibitem[\protect\citeauthoryear{{Nidever} et~al.,}{{Nidever}
  et~al.}{2015}]{nidever+15}
{Nidever} D.~L.,  et~al., 2015, \mn@doi [\aj] {10.1088/0004-6256/150/6/173},
  \href {http://adsabs.harvard.edu/abs/2015AJ....150..173N} {150, 173}

\bibitem[\protect\citeauthoryear{{Nidever} et~al.,}{{Nidever}
  et~al.}{2019}]{nidever+19}
{Nidever} D.~L.,  et~al., 2019, arXiv e-prints, \href
  {https://ui.adsabs.harvard.edu/\#abs/2019arXiv190103448N} {p.
  arXiv:1901.03448}

\bibitem[\protect\citeauthoryear{{Nissen} \& {Schuster}}{{Nissen} \&
  {Schuster}}{2010}]{nissen+10}
{Nissen} P.~E.,  {Schuster} W.~J.,  2010, \mn@doi [\aap]
  {10.1051/0004-6361/200913877}, \href
  {https://ui.adsabs.harvard.edu/#abs/2010A&A...511L..10N} {511, L10}

\bibitem[\protect\citeauthoryear{{Nomoto}, {Kobayashi}  \& {Tominaga}}{{Nomoto}
  et~al.}{2013}]{nomoto+13}
{Nomoto} K.,  {Kobayashi} C.,   {Tominaga} N.,  2013, \mn@doi [Annual Review of
  Astronomy and Astrophysics] {10.1146/annurev-astro-082812-140956}, \href
  {https://ui.adsabs.harvard.edu/\#abs/2013ARA&A..51..457N} {51, 457}

\bibitem[\protect\citeauthoryear{{Piffl} et~al.,}{{Piffl}
  et~al.}{2014}]{piffl+14}
{Piffl} T.,  et~al., 2014, \mn@doi [\mnras] {10.1093/mnras/stu1948}, 445, 3133

\bibitem[\protect\citeauthoryear{{Pillepich}, {Madau}  \& {Mayer}}{{Pillepich}
  et~al.}{2015}]{pillepich+15}
{Pillepich} A.,  {Madau} P.,   {Mayer} L.,  2015, \mn@doi [\apj]
  {10.1088/0004-637X/799/2/184}, \href
  {https://ui.adsabs.harvard.edu/\#abs/2015ApJ...799..184P} {799, 184}

\bibitem[\protect\citeauthoryear{{Read}, {Iorio}, {Agertz}  \&
  {Fraternali}}{{Read} et~al.}{2017}]{read+17}
{Read} J.~I.,  {Iorio} G.,  {Agertz} O.,   {Fraternali} F.,  2017, \mn@doi
  [\mnras] {10.1093/mnras/stx147}, \href
  {https://ui.adsabs.harvard.edu/\#abs/2017MNRAS.467.2019R} {467, 2019}

\bibitem[\protect\citeauthoryear{{Revaz} \& {Jablonka}}{{Revaz} \&
  {Jablonka}}{2012}]{revaz+12}
{Revaz} Y.,  {Jablonka} P.,  2012, \mn@doi [\aap]
  {10.1051/0004-6361/201117402}, \href
  {https://ui.adsabs.harvard.edu/\#abs/2012A&A...538A..82R} {538, A82}

\bibitem[\protect\citeauthoryear{{Robin}, {Marshall}, {Schultheis}  \&
  {Reyl{\'e}}}{{Robin} et~al.}{2012}]{robin+12}
{Robin} A.~C.,  {Marshall} D.~J.,  {Schultheis} M.,   {Reyl{\'e}} C.,  2012,
  \mn@doi [\aap] {10.1051/0004-6361/201116512}, \href
  {https://ui.adsabs.harvard.edu/#abs/2012A&A...538A.106R} {538, A106}

\bibitem[\protect\citeauthoryear{{Rodr{\'\i}guez-Puebla}, {Avila-Reese}  \&
  {Drory}}{{Rodr{\'\i}guez-Puebla} et~al.}{2013}]{rodriguezpuebla+13}
{Rodr{\'\i}guez-Puebla} A.,  {Avila-Reese} V.,   {Drory} N.,  2013, \mn@doi
  [\apj] {10.1088/0004-637X/773/2/172}, \href
  {https://ui.adsabs.harvard.edu/\#abs/2013ApJ...773..172R} {773, 172}

\bibitem[\protect\citeauthoryear{{Sanders} \& {Das}}{{Sanders} \&
  {Das}}{2018}]{sanders+18}
{Sanders} J.~L.,  {Das} P.,  2018, \mn@doi [\mnras] {10.1093/mnras/sty2490},
  \href {https://ui.adsabs.harvard.edu/\#abs/2018MNRAS.481.4093S} {481, 4093}

\bibitem[\protect\citeauthoryear{{Sch{\"o}nrich}, {Asplund}  \&
  {Casagrande}}{{Sch{\"o}nrich} et~al.}{2011}]{schonrich+11}
{Sch{\"o}nrich} R.,  {Asplund} M.,   {Casagrande} L.,  2011, \mn@doi [\mnras]
  {10.1111/j.1365-2966.2011.19003.x}, \href
  {http://adsabs.harvard.edu/abs/2011MNRAS.415.3807S} {415, 3807}

\bibitem[\protect\citeauthoryear{{Sch{\"o}nrich}, {Asplund}  \&
  {Casagrande}}{{Sch{\"o}nrich} et~al.}{2014}]{schonrich+14}
{Sch{\"o}nrich} R.,  {Asplund} M.,   {Casagrande} L.,  2014, \mn@doi [\apj]
  {10.1088/0004-637X/786/1/7}, \href
  {http://adsabs.harvard.edu/abs/2014ApJ...786....7S} {786, 7}

\bibitem[\protect\citeauthoryear{{Schuster}, {Moreno}, {Nissen}  \&
  {Pichardo}}{{Schuster} et~al.}{2012}]{schuster+12}
{Schuster} W.~J.,  {Moreno} E.,  {Nissen} P.~E.,   {Pichardo} B.,  2012,
  \mn@doi [\aap] {10.1051/0004-6361/201118035}, 538, A21

\bibitem[\protect\citeauthoryear{{Sesar}, {Juri{\'c}}  \& {Ivezi{\'c}}}{{Sesar}
  et~al.}{2011}]{sesar+11}
{Sesar} B.,  {Juri{\'c}} M.,   {Ivezi{\'c}} {\v Z}.,  2011, \mn@doi [\apj]
  {10.1088/0004-637X/731/1/4}, 731, 4

\bibitem[\protect\citeauthoryear{{Sesar} et~al.,}{{Sesar}
  et~al.}{2013}]{sesar+13}
{Sesar} B.,  et~al., 2013, \mn@doi [\aj] {10.1088/0004-6256/146/2/21}, \href
  {http://adsabs.harvard.edu/abs/2013AJ....146...21S} {146, 21}

\bibitem[\protect\citeauthoryear{{Shetrone} et~al.,}{{Shetrone}
  et~al.}{2015}]{shetrone+15}
{Shetrone} M.,  et~al., 2015, \mn@doi [The Astrophysical Journal Supplement
  Series] {10.1088/0067-0049/221/2/24}, \href
  {https://ui.adsabs.harvard.edu/\#abs/2015ApJS..221...24S} {221, 24}

\bibitem[\protect\citeauthoryear{{Shetrone} et~al.,}{{Shetrone}
  et~al.}{2019}]{shetrone+19}
{Shetrone} M.,  et~al., 2019, arXiv e-prints, \href
  {https://ui.adsabs.harvard.edu/\#abs/2019arXiv190109592S} {p.
  arXiv:1901.09592}

\bibitem[\protect\citeauthoryear{{Sirko} et~al.,}{{Sirko}
  et~al.}{2004}]{sirko+04}
{Sirko} E.,  et~al., 2004, \mn@doi [\aj] {10.1086/381486}, \href
  {http://adsabs.harvard.edu/abs/2004AJ....127..914S} {127, 914}

\bibitem[\protect\citeauthoryear{{Smith} et~al.,}{{Smith}
  et~al.}{2009}]{smith+09b}
{Smith} M.~C.,  et~al., 2009, \mn@doi [\mnras]
  {10.1111/j.1365-2966.2009.15391.x}, \href
  {http://adsabs.harvard.edu/abs/2009MNRAS.399.1223S} {399, 1223}

\bibitem[\protect\citeauthoryear{{Tissera}, {White}  \&
  {Scannapieco}}{{Tissera} et~al.}{2012}]{tissera+12}
{Tissera} P.~B.,  {White} S. D.~M.,   {Scannapieco} C.,  2012, \mn@doi [\mnras]
  {10.1111/j.1365-2966.2011.20028.x}, \href
  {https://ui.adsabs.harvard.edu/\#abs/2012MNRAS.420..255T} {420, 255}

\bibitem[\protect\citeauthoryear{{Vasiliev}}{{Vasiliev}}{2019}]{vasiliev+19}
{Vasiliev} E.,  2019, \mn@doi [\mnras] {10.1093/mnras/sty2672}, \href
  {https://ui.adsabs.harvard.edu/\#abs/2019MNRAS.482.1525V} {482, 1525}

\bibitem[\protect\citeauthoryear{{Vincenzo}, {Spitoni}, {Calura}, {Matteucci},
  {Aguirre}, {Miglio}  \& {Cescutti}}{{Vincenzo} et~al.}{2019}]{vincenzo+19}
{Vincenzo} F.,  {Spitoni} E.,  {Calura} F.,  {Matteucci} F.,  {Aguirre} V.~S.,
  {Miglio} A.,   {Cescutti} G.,  2019, arXiv e-prints, \href
  {https://ui.adsabs.harvard.edu/\#abs/2019arXiv190303465V} {p.
  arXiv:1903.03465}

\bibitem[\protect\citeauthoryear{{Watkins} et~al.,}{{Watkins}
  et~al.}{2009}]{watkins+09}
{Watkins} L.~L.,  et~al., 2009, \mn@doi [\mnras]
  {10.1111/j.1365-2966.2009.15242.x}, 398, 1757

\bibitem[\protect\citeauthoryear{{Williams} \& {Evans}}{{Williams} \&
  {Evans}}{2015}]{williams+15b}
{Williams} A.~A.,  {Evans} N.~W.,  2015, \mn@doi [\mnras]
  {10.1093/mnras/stv1967}, \href
  {http://adsabs.harvard.edu/abs/2015MNRAS.454..698W} {454, 698}

\bibitem[\protect\citeauthoryear{{Xue}, {Rix}, {Ma}, {Morrison}, {Bovy},
  {Sesar}  \& {Janesh}}{{Xue} et~al.}{2015}]{xue+15}
{Xue} X.-X.,  {Rix} H.-W.,  {Ma} Z.,  {Morrison} H.,  {Bovy} J.,  {Sesar} B.,
  {Janesh} W.,  2015, preprint (\mn@eprint {arXiv} {1506.06144})

\bibitem[\protect\citeauthoryear{{Yong} et~al.,}{{Yong} et~al.}{2016}]{yong+16}
{Yong} D.,  et~al., 2016, \mn@doi [\mnras] {10.1093/mnras/stw676}, \href
  {http://adsabs.harvard.edu/abs/2016MNRAS.459..487Y} {459, 487}

\bibitem[\protect\citeauthoryear{{Zamora} et~al.,}{{Zamora}
  et~al.}{2015}]{zamora+15}
{Zamora} O.,  et~al., 2015, \mn@doi [\aj] {10.1088/0004-6256/149/6/181}, \href
  {https://ui.adsabs.harvard.edu/\#abs/2015AJ....149..181} {149, 181}

\bibitem[\protect\citeauthoryear{{Zolotov}, {Willman}, {Brooks}, {Governato},
  {Brook}, {Hogg}, {Quinn}  \& {Stinson}}{{Zolotov} et~al.}{2009}]{zolotov+09}
{Zolotov} A.,  {Willman} B.,  {Brooks} A.~M.,  {Governato} F.,  {Brook} C.~B.,
  {Hogg} D.~W.,  {Quinn} T.,   {Stinson} G.,  2009, \mn@doi [\apj]
  {10.1088/0004-637X/702/2/1058}, \href
  {https://ui.adsabs.harvard.edu/\#abs/2009ApJ...702.1058Z} {702, 1058}

\bibitem[\protect\citeauthoryear{{de Jong}, {Yanny}, {Rix}, {Dolphin}, {Martin}
   \& {Beers}}{{de Jong} et~al.}{2010}]{dejong+10}
{de Jong} J.~T.~A.,  {Yanny} B.,  {Rix} H.-W.,  {Dolphin} A.~E.,  {Martin}
  N.~F.,   {Beers} T.~C.,  2010, \mn@doi [\apj] {10.1088/0004-637X/714/1/663},
  714, 663

\makeatother
\end{thebibliography}



\appendix


\bsp	
\label{lastpage}
\end{document}